# Physically Informed Bayesian Retrieval of SWE and Snow Depth in Forested Areas from Airborne X And Ku-Band SAR Measurements

Siddharth Singh[1], Carrie Vuyovich[2] Ana P. Barros[1]

[1]Department of Civil and Environmental Engineering, University of Illinois at Urbana-Champaign, Urbana, Illinois, USA
[2]NASA Goddard Space Flight Center - Greenbelt, MD

*Correspondence to*: Ana P. Barros (barros@illinois.edu)

**Abstract**

This study presents a coupled atmospheric physical-statistical Snow Water Equivalent (SWE) retrieval framework in forested areas using dual-frequency X- and Ku-band SAR measurements. The methodology builds on previous work coupling snow hydrology and microwave propagation and backscatter models and introduces a parameterization of microwave propagation and scattering the forest canopy based on the Water Cloud Model (WCM) modified to account for canopy closure effects. The retrieval framework was applied to airborne SnowSAR measurements over Grand Mesa, Colorado and performance was evaluated against snow pit observations and LiDAR snow depth estimates. Prior distributions of snowpack properties were generated using a multilayer snow hydrology model (MSHM) forced by Numerical Weather Prediction (NWP) forecasts. Prior distributions of vegetation and ground parameters were initialized using Ku-HH measurements, with effective soil and vegetation parameters estimated under frozen conditions. Ground parameters were estimated in open areas and spatially interpolated to nearby forested areas using ordinary kriging. Successful SWE and snow depth retrievals are achieved for forested pixels with relative backscatter tolerance in the Bayesian optimization below 30% for pixels and incidence angles between 30°–50° along SnowSAR flight paths. Successful retrievals capture both the mean and variance of snowpack distributions across the Grand Mesa plateau. Validation against collocated LiDAR snow depth and snow pit SWE measurements from the SnowEx'17 campaign showed a root mean square error (RMSE) of 0.033 m (< 8% of maximum SWE for pits) for forested pixels at 90 m spatial resolution, with improved agreement in spatial patterns compared to snow hydrology predictions driven by NWP alone. The performance deteriorates over heterogenous land-cover (e.g. mixed forest and wetlands) at subpixel scale such as forest boundaries and land-margins of ponds and lakes due to uncertainty in the estimation of canopy closure. Nevertheless, absolute differences between LiDAR snow depth and retrievals less than 10% and 20% for 62% and 82% pixels, respectively. Retrievals at 30 m resolution for one flight demonstrated reduced spatial errors in heterogeneous terrain, with an increase in the retrieval efficiency by 78% (for 10% absolute relative error). The study demonstrates the feasibility of dual-frequency Bayesian SWE retrieval algorithm in forested landscapes by combining physical modeling with remote sensing.

# 1. Introduction

Snow accumulation changes in cold regions such as the Arctic are monitored with great interest as they reflect concerted changes in precipitation patterns as well as regional weather (Lee et al., 2021; Curk et al., 2020; Shi and Liu, 2021; Switanek et al., 2024; Kacimi and Kwok, 2022; Pongracz et al., 2024;Gottlieb and Mankin, 2024). Snow cover and snowpack microphysical properties govern terrestrial albedo over large regions of the world and thus play a key role in regulating the planet's energy budget (Fassnacht et al., 2016; Xu and Dirmeyer, 2013; Jennings and Molotch, 2020). Snowpacks represent an important form of transient water storage in the cold season (Rodell and Houser, 2004; Lim et al., 2021; Mazzotti et al., 2024) followed by melt and runoff in the warm season.  Monitoring and predicting snowpack properties is essential for a myriad of applications from water resources management to agriculture production to flood response and mitigation (Gardner et al., 2013; Semádeni-Davies, 2004; Falk and Lin, 2019; Nicolaus et al., 2021; Sthapit et al., 2022; Horrigan and Bates, 1995; Li et al., 2025).

Remote sensing of snowpack properties relies on Mie scattering theory, which describes the interaction between electromagnetic waves (EMW) and particles based on their relative sizes (Aoki et al., 2000; Hall et al., 2004; Tsang et al., 2007).  Thus, the EM wavelength and the diameter of the particle (D) determine the type of information gathered. When the EM wavelength ($\lambda$) is much smaller than the particle diameter D ($\lambda \ll D$), surface reflectance is the dominant scattering process. Conversely, when the wavelength is equal to or larger than the particle diameter ($\lambda \geq D$), volumetric scattering is dominant, revealing information about the internal structure of the snowpack. The trade-off between surface and volume scattering is crucial in selecting and combining appropriate wavelengths for remote sensing applications. Previous research has demonstrated value in the combination of X and Ku-band backscatter measurements to quantify snow mass properties from volume backscatter (Singh et al., 2024 (a), Boyd et al., 2022, Tsang et al., 2021)

Snow water equivalent (SWE), obtained as the product of snow depth and snow density, represents the amount of water stored in a snowpack if melted completely. Thus, to estimate SWE is to estimate snow water resources. Statistical models that integrate SWE estimates from microwave backscatter observations with ground-based measurements, often through cost minimization or data assimilation approaches, have been widely employed to estimate SWE at high spatial resolution (Mote et al. 2005; Li et al. 2017; Zhu et al., 2021). However, purely statistical or machine-learning approaches frequently show reduced accuracy when extrapolated to higher-resolution or more heterogeneous datasets, where scale and physical complexity differ from the training domain (Bonavita 2024; Hernanz et al., 2024; Slater et al., 2025). This issue will become more important with the increasing availability of high-resolution remote sensing data (Wrzesien et al., 2017; Sabetghadam et al., 2025; Boueshagh et al., 2025). Data assimilation into snow hydrology models provides a general path for SWE estimation constrained by physical principles (Sturm et al., 2010; Shrestha and Barros 2025).

Cao and Barros (2020) integrated the Multilayered Snow Hydrology Model (MSHM) earlier developed by Kang and Barros (2011 a and b) and simulates the temporal evolution of snowpacks and captures detailed changes in snow stratigraphy and internal structure with the Microwave

Emission Model of Multilayered Snowpacks (MEMLS, Proksch et al., 2015) for forward simulations of snowpack microstructure. Pan et al. (2023) implemented MEMLS in a Bayesian framework, referred to as BASE-AM, to estimate SWE from active microwave backscatter measurements. Building on these, Singh et al. (2024) modified the BASE-AM algorithm to derive snowpack priors from MSHM simulations driven by weather forecasts and to improve ground backscatter estimates for frozen soils. They applied the modified algorithm to retrieve SWE from Ku- and X-band SnowSAR observations from the NASA SnowEx'17 campaign in Grand Mesa, Colorado achieving an RMSE of less than 7% when compared with snow pit observations in open snow-covered grasslands.

The scattering behavior of active microwaves in forested snowpacks is very complex including interactions among vegetation, snowpack, submerged vegetation, and ground (Figure 1; Mahat and Tarboton, 2012; Essery et al., 2024) resulting in strong attenuation of backscatter and challenging separation of scattering and attenuation sources (Cho et al., 2022; Lemmetyinen et al. 2022). The goal of this study is to extend the physical-statistical retrieval framework from Singh et al. (2024) to forested environments. Retrieving SWE in forested landscapes remains a major challenge for snow remote sensing, yet it is essential because forests account for roughly one-third of Earth's seasonal snow-covered area (Bonnell et al., 2024)

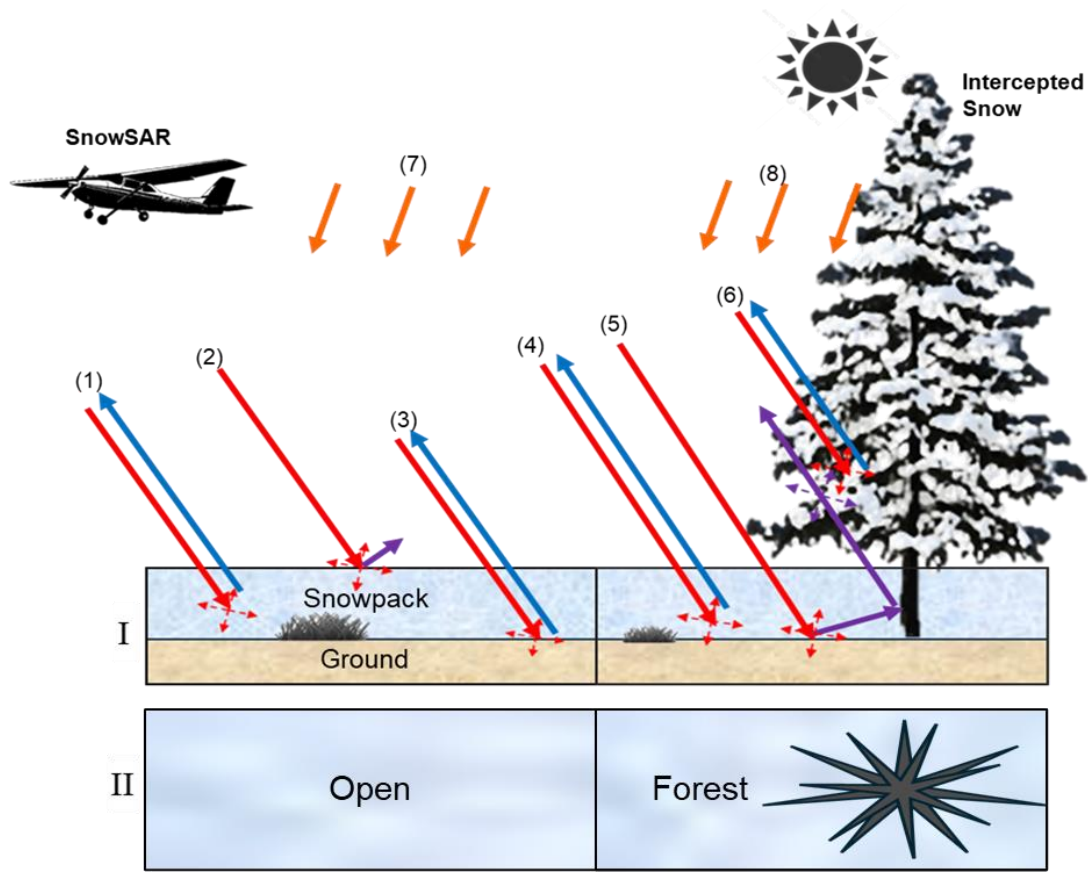

**Figure 1** I) Side view of scattering mechanisms for neighboring grassland and forest pixels submerged by snow and snowpack over bare soil or rock: (1) Volume Backscatter $\sigma_{vol}$; (2) surface backscatter $\sigma_{surf}$; (3) background backscatter at the snow-ground interface $\sigma_{bkg}$; (4) Volume Backscatter $\sigma_{vol}$ not affected by vegetation in a forest designated pixel (5) snowpack-ground-canopy-

tree interaction with one way vegetation backscatter (6) Two way vegetation backscatter σ<sub>veg</sub> at high frequency (7) Incoming solar radiation in open area (8) Solar radiation in forest pixel requiring correction. II) Top view of pixels with illustrating canopy closure.

## 1.1. Parameterization of Vegetation Backscatter

Simulation of vegetation backscatter depends on factors such as vegetation type, leaf orientation and distribution, vegetation temperature, and water content. Vegetation backscatter models are typically categorized into three types: (i) empirical, (ii) semi-empirical, and (iii) theoretical. Empirical models, such as the Dubois model and machine learning models , rely heavily on large, site-specific datasets to estimate radar backscatter (Dubois et al., 1995; Baghdadi et al., 2012; El Hajj et al., 2016; Mueller et al., 2022). As a result, these models are not well-suited for use in radiative transfer-based retrieval algorithms for general applications. Semiempirical models such as the Oh and the Water Cloud Model (WCM) integrate site specific parameters within a physical framework (Attema and Ulaby, 1978; Bindlish and Barros, 2000; Li and Wang, 2018; Oh et al., 2022). In contrast, theoretical models like the Integral Equation Model (IEM) use detailed vegetation properties including canopy structure, water content, and biomass to simulate radar backscatter (Bindlish and Barros, 2000; Khabazan et al., 2013; Panciera et al., 2013). However, the extensive input requirements of these latter models limit their applicability in large-scale or operational retrievals, where such detailed vegetation characteristics and other ancillary data are often unavailable. These challenges highlight the need for scalable, physically grounded models that can balance complexity with data availability to improve vegetation correction in snowpack retrievals.

The WCM combines transmissivity, vegetation backscatter and ground backscatter parameters to determine total backscatter from a vegetated pixel (Attema and Ulaby 1978; Bindlish and Barros 2001; Vermunt et al., 2022). Previous WCM applications have been directed at soil moisture retrievals from radar measurements. In a typical WCM application the vegetation backscatter is estimated using empirical relations based on ancillary data and or remote-sensing vegetation indices and optimization of frequency and polarization dependent parameters against observations (Bindlish and Barros, 2001; Kumar et al., 2012, Li and Wang (2018), Park et al., 2019, Qin et al., 2024). Here, the WCM is adapted to represent vegetation effects in SAR measurements over snow covered forested areas as follows.

First, the total backscatter $\sigma_{tot}$ includes contributions of vegetation backscatter from the forest canopy $\sigma_{veg}$, the volume backscatter from the snowpack and at the snow-ground interface (background) including both soil and submerged vegetation $\sigma_{vol+bkg}$ transmitted by the canopy, and the vegetation backscatter $\sigma_{veg}$ (Figure 1):

$$\sigma_{total} = \sigma_{veg} + \tau_{veg}(\sigma_{vol+bkg}) \quad (1)$$

$$\sigma_{vol+bkg} = \sigma_{vol} + \tau_{vol}\sigma_{bkg} \quad (2)$$

$$\sigma_{veg} = AM_v cos\theta \quad (3)$$

where $M_v$ is vegetation water content, and $\theta$ is the viewing angle. The transmissivity varies exponentially with the vegetation water content

$$\tau_{veg} = e^{-2BM_v sec\theta} \quad (4)$$

Here A and B site specific, frequency and polarization dependent calibration parameters.

## 1.2. Radiation in Forested Environment

In addition to scattering, vegetation contributes to attenuation of solar radiation via absorption in the canopy (Cao and Barros 2023, Hardy et al., 2004, Wang et al., 2007, Essery et al., 2008). Most canopy radiative transfer schemes parameterize transmissivity (τ) using the Beer–Lambert law or two-stream approaches (Myneni et al., 1989; Sellers, 1985, 1987), quantifying the fraction of radiation that penetrates the canopy after interacting with foliage through absorption and scattering. However, transmissivity alone does not capture the spatial variability in canopy density. During winter, many deciduous trees lose their leaves, leading to substantial reductions in leaf area index (LAI) and canopy closure (Wang et al., 2016).

Because the degree of attenuation depends not only on the optical properties of the canopy but also on the fraction of sky obscured by vegetation, incorporating canopy closure ($C_c$) into the radiation transmissivity formulation is essential for accurately modifying the radiation incident upon the snowpack to account for shading and attenuation effects. The net downward flux of radiation reaching the snow surface beneath a canopy can be expressed as the sum of two components: (i) radiation transmitted directly through the open sky fraction (1−$C_c$), and (ii) radiation attenuated as it goes through the canopy. This leads to the following expressions for shortwave and longwave radiation beneath the canopy:

$$SW_{corr} = (1 - C_c)SW + \tau_{sw}C_c SW \quad (5)$$

$$LW_{corr} = (1 - C_c)LW + C_c(\tau_{lw}LW + \kappa\epsilon_v T_{veg}^4) \quad (6)$$

Here, the first term in each equation represents the fraction of radiation that reaches the snowpack directly through canopy openings, while the second term accounts for radiation transmitted through and emitted by the canopy layer. Assuming full-canopy conditions with minimal canopy gap fractions, the transmissivity terms ($\tau_{sw}$ and $\tau_{lw}$) reflect the effective attenuation within dense forest stands with large observed canopy closure. Specifically, shortwave transmissivity ($\tau_{sw}$) for conifer trees (Hardy et al. 2004) and decidous trees (Hardy et al., 1998); Longwave transmissivity ($\tau_{lw}$) is derived from Asner et al. (1998) and longwave emissivity ($\epsilon_v$) is obtained from Engineering Toolbox (2003).

A generalized model for remotely estimating canopy closure remains unavailable. Canopy closure ($C_c$) has been estimated using empirical models derived from field observations that relate measurable structural and spectral properties such as leaf area index (LAI), stand density, and gap fraction to closure (Danson et al., 2007; Seidel et al., 2016). Many of these empirical approaches are developed for specific canopy types and often rely on vegetation reflectance–based parameters as a proxy for closure. For example, Pomeroy et al. (2002) proposed an empirical parameterization canopy closure of conifer trees to Leaf Area Index (LAI):

$$C_c = 0.29 \log(LAI) + 0.55 \tag{7}$$

The Leaf Area Index (LAI) is estimated from satellite-based surface reflectance. Rasmus et al. (2013) showed that while the empirical relationship between LAI and canopy closure in Eq. (7) works well for uniform forest stands with similar tree density and structure, its accuracy decreases in heterogeneous forests where canopy gaps and sub-canopy light conditions vary greatly. These uncertainties propagated into snow models, affect snowmelt timing and energy balance, particularly during the ablation period, and must be carefully assessed, particularly when applying the model across diverse forest types at varying spatial resolutions. Therefore, improvements in canopy structure estimation will be essential for enhancing the accuracy of snowpack retrievals in forested pixels. Bindlish and Barros (2001) introduced parameter dependence on vegetation architecture to distinguish among different types of crops in their application of the WCM parameterization. Given time-series of measurements such as those available from satellite missions revisits, it is possible to introduce such dependencies for forests with mixed tree species as the repeated measurements enable pixel (site specific) parameter estimation. In the absence of field measurements given the limited number of overlapping SnowSAR flights for Grand Mesa, and the lack of alternative parameterizations for different tree species and tree architectures, the Pomeroy et al. (2002) approach is adopted here. Both the field site in Pomeroy et al. (2002) and Grand Mesa are conifer-dominated, snow-bearing forests with similar canopy architecture and cold continental climates, and thus Eq.(7) provides a reasonable approximation of canopy closure for this study.

### 1.3. Precipitation and Interception

Precipitation within forested environments should be modified according to interception efficiency of the overlying canopy. Fresh precipitation is reduced due to interception of snowfall in the canopy. High density is added to the snowpack due to unloading of intercepted snowfall. Multiple empirical models have been developed to determine interception in boreal forests (Hedstorm and Pomeroy, 1998; Lundquist et al., 2021; Helbig et al., 2020). Tree architecture and stand density and diversity of tree species all impact interception. Stork et al. (2002) installed lysimeters to measure the mass balance on and around four trees and estimated a factor of 0.6 for interception. Hedstorm and Pomeroy (1998) derived a semi-empirical interception model fitting field measurements that is widely used:

$$I_{max} = 6.3 LAI \left(0.27 + \frac{46}{\rho}\right) \tag{8}$$

$$dI_i = 0.68(I_{max} - I_{i-1})\left(1 - e^{-\frac{P_i C_c}{I_{max}}}\right) \tag{9}$$

$$P_{i,corr} = P_i - dI_i \tag{10}$$

$$I_i = I_{i-1} + dI_i - f_u - f_m \tag{11}$$

Here $I_{max}$ is the maximum canopy interception; $P_i$ is the snowfall and $\rho$ is fresh snow density that is set as 30 kg/m³ in the model (Cao and Barros 2023; Singh et al., 2024); $f_u$ and $f_m$ are wind and melt unloading rates, respectively. Wind unloading can be quantified following Roesch et al. (2001):

$$fu = 0.0231W \tag{12}$$

where $W$ is the average wind speed in ms⁻¹. Melt unloading and sublimation of intercepted snow are not considered in accumulation season simulations presented here. The unloaded snow is added to the snowpack in MSHM. The density of unloaded snow is estimated following Bouchard et al. (2022):

$$\rho_{s,int} = \rho_{fr} + (\rho_{max} - \rho_{fr})\left(1 - e^{-\frac{a_{s,int}}{\tau}}\right) \tag{13}$$

Here $\rho_{s,int}$ is the density of intercepted snow, $\rho_{fr}$ is density of fresh snow taken as 30 kg/m³. $\tau$ is an independent parameter such that $\rho_{s,int}$ reaches 99% of the maximum density of intercepted snow, $a_{s,int}$ within 30 days. If new snow is intercepted within 30 days, $a_{s,int}$ changes to,

$$a_{s,int} = a_{s,int}\left(1 - \frac{\Delta I}{I + \Delta I \Delta t}\right) \tag{14}$$

where $\Delta t$ is computational period, 30 min in our case. Note that these empirical parameterizations of interception are site specific and their transferability has not been rigorously assessed to quantify uncertainty. Parameters such as interception efficiency and unloading coefficients are calibrated at point scale and may not generalize well to entire forest stands and larger spatial scales, limiting the robustness of SWE estimates in heterogeneous forested environments. Key sources to uncertainty include variability in canopy structure across forest types, inaccurate or generalized LAI and canopy closure inputs, errors in estimating wind speed at canopy level; simplifications in unloading dynamics, and uncertainty in fresh snow density, which may vary significantly with meteorological conditions.

## 2. Study Area and Datasets

The study is conducted over Grand Mesa, Colorado, a high-elevation plateau situated approximately 2,000 m above surrounding lowlands and bordered by ridges rising to 500 m (Singh et al., 2024). Grand Mesa experiences an alpine climate with persistent snowfall outside of July and August. The region exhibits heterogeneous land cover, with grasslands predominantly in the west and a mix of evergreen and deciduous forests toward the east, interspersed with numerous wetlands, especially across ecotonal zones. Land cover classification is based on the National Land Cover Dataset (NLCD) and resampled to 90 m using nearest neighbor interpolation to support retrievals at that scale, consistent with the methodology described in Singh et al. (2024). Hourly albedo is derived from 12.5 km National Land Data Assimilation system (NLDAS) fields. Canopy closure is estimated using MODIS LAI data (MOD15A2H) in combination with the Global Land Analysis and Discovery (GLAD) tree height dataset (Myneni et al., 2015; Potapov et al., 2020).

LAI is downscaled from 500 m to 30 m using Global Tree height as a proxy as shown in Appendix A. Table 1 provides a comprehensive summary of the datasets, raw and final resolution, sensors used to acquire the datasets and access links, as used in the SWE retrieval algorithm.

**Table 1-** Summary list of datasets used in the study

| Data | Source/Sensor | Spatial Resolution | | Temporal Resolution | | Date Range | Relevant Link |
|---|---|---|---|---|---|---|---|
| | | Initial | Final | Initial | Final | | |
| Rainfall Temperature Air Pressure Incoming SW radiation Incoming Longwave radiation Wind speed Humidity | HRRR | 3 km | 90 m | 1 hr | 30 min | 9/1/2016 - 2/25/2017 | https://rapidrefresh.noaa.gov/hrrr/ |
| Albedo | NLDAS | 12.5 km | 30 m | 1 hr | 30 min | 9/1/2016 - 2/25/2017 | https://ldas.gsfc.nasa.gov/ |
| Backscatter | SnowSAR – SnowEx'17 | 1 m | 90 m | - | - | 2/21/2017 | https://nsidc.org/data/snex17_snowsar/versions/1 |
| Landcover | NLCD | 30 m | 90 m | - | - | - | https://www.usgs.gov/centers/eros/science/national-land-cover-database |
| Snow Depth | LiDAR – SnowEx'17 | 3 m | 90 m | - | - | 2/25/2017 | https://nsidc.org/data/aso_3m_sd/versions/1 |
| SWE | Snowpit – SnowEx'17 | - | - | - | - | 2/20/2017 - 2/24/2017 | https://nsidc.org/data/snex17_snowpits/versions/1 |
| LAI | MODIS | 500 m | 90 m | 1 Day | 1 Day | 9/1/2016 - 2/25/2017 | https://modis.gsfc.nasa.gov/data/dataprod/mod15.php |
| Tree Height | GLAD | 30 m | 90 m | - | - | 9/1/2016 - 2/25/2017 | https://glad.umd.edu/dataset/gedi |

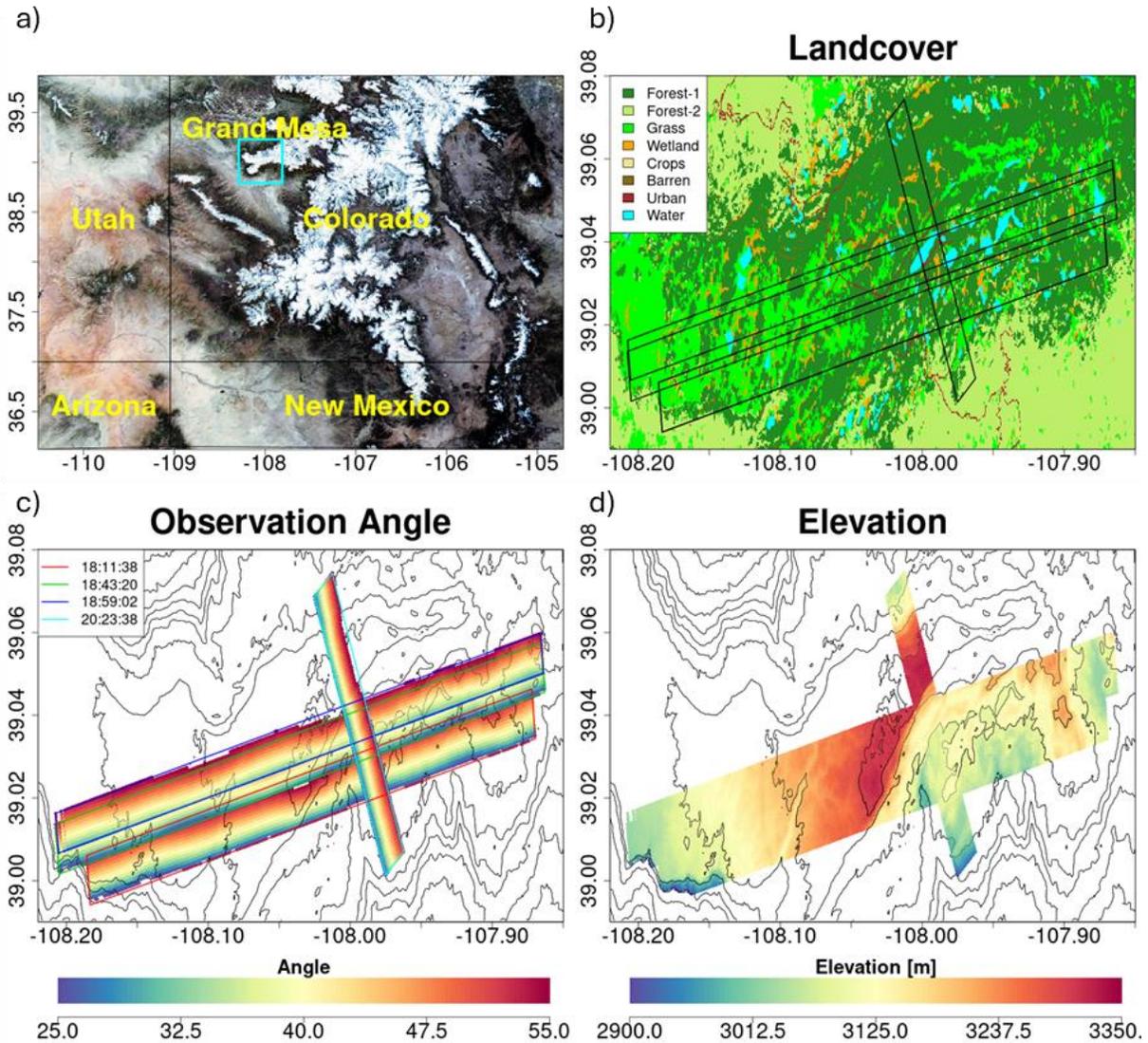

**Figure 2** - Study area in Grand Mesa, Colorado. a) Location of Grand Mesa in Colorado, with historical Apr 1 SWE average as base map. . b) Land cover of the study region. Forest-1 are needle leaf forests; Forest-2 are broadleaf forests. c) Paths of 4 SnowSAR SnowEx'17 flights on 21 Feb 2017, with true color image obtained from Landsat on 03/11/2017 as the base map d) Digital elevation map of the study region.

## 2.1 Atmospheric Forcing

Numerical Weather Prediction (NWP) data are used to provide atmospheric forcing and boundary conditions for the snow hydrology model. Following the approach in Singh et al. (2024), we utilize hourly forecasts from the High-Resolution Rapid Refresh (HRRR) model developed by the National Oceanic and Atmospheric Administration (NOAA), which assimilates observations at 3 km resolution (Benjamin et al., 2016). We modified the incoming shortwave and longwave radiation from HRRR using Equation 5 for canopy shadow effects in forested areas. Canopy closure ($C_c$) is derived by downscaling the MODIS LAI dataset (500 m resolution) to 30 m, using tree height as a covariate (See Appendix 1). Snowfall interception is calculated using equation (7) and subtracted from the incoming precipitation, while the unloading of intercepted snow is subsequently added to the snowpack. The density of unloaded snow, which differs from fresh snow due to metamorphic processes, is calculated using equation (13). Atmospheric forcing variables are linearly interpolated to match the 30-minute temporal resolution 90 m spatial resolution to be used in the model.

## 2.2 SnowSAR Backscatter

Airborne microwave backscatter measurements were acquired over Grand Mesa on 21 February 2017 during the NASA SnowEx campaign using the SnowSAR instrument, a dual-frequency (X- and Ku-band) synthetic aperture radar system (see Singh et al., 2024; Table 1). The data were collected at ~1 m resolution along six flightlines - two over steep, densely forested terrain and four over the plateau. Only the plateau flightlines are used in this study (Fig. 2, Fig. 3), corresponding to flight times between 18:00 and 21:00 GMT (12:00–15:00 MST). SnowSAR data underwent rigorous quality control, including filtering based on aircraft attitude (e.g., turbulence-induced instability), beam incidence angle, antenna pattern, and signal-to-noise ratio. As established in prior analyses, only the co-polarized (HH and VV) backscatter measurements are retained for retrieval due to their consistent signal quality. Geolocation accuracy was validated using corner reflectors and prominent geographic features. Figure 2 shows the flight paths and the observation angles of the retrieved backscatter measurements. Figure 3 shows the distribution of observed backscatter for open and forested areas. A large difference in backscatter is measured for X-HH compared to X-VV, Ku-VV and Ku-HH, which increases in the forested areas.  While the X-HH measurements are used to estimate ground parameters similar to Singh et al. (2024, Step II in Fig. 4), only Ku-HH measurements are used to estimate vegetation parameters for the WCM parameterization for retrievals (Step III, Fig. 4) to achieve faster convergence in the final step of Bayesian optimization to retrieve SWE using X-VV and Ku-VV (Step IV, Fig. 4).  Further details are reviewed in Section 3. The original 1 m SnowSAR data was aggregated to 90 m grids by averaging all valid observations within each pixel.

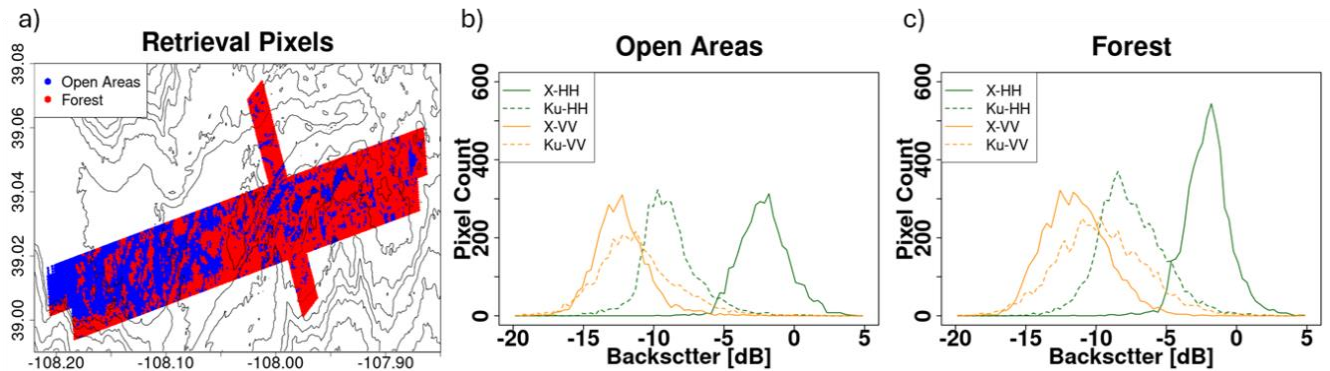

**Figure 3** – a) Spatial distribution **of** retrieved pixels at 90 m resolution differentiating between open (blue) and forested (red) areas along the flight path lines. Singh et al. (2024) retrievals were for the open areas. This study proposes a retrieval algorithm to estimate snowpack properties in the forested region.  The distributions of SnowSAR backscatter for X and Ku Band, HH and VV polarizations are shown in (b) for open areas and in  (c) for forested pixels.

## 2.3 Validation Data

*LiDAR Snow Depth*: Airborne Snow Observatory (ASO) LiDAR measurements of snow depth at 3 m spatial resolution were acquired over Grand Mesa on 25 February 2017, four days after the SnowSAR flights (Painter et al., 2018; Table 1). No significant snowfall or high-wind events occurred during this period, aside from a brief (~1 hour) rainfall event of approximately 3 mm on 24 February. The ASO LiDAR data are used to assess the spatial distribution of retrieved snow depths, characterizing snowpack heterogeneity, and to quantify absolute differences between retrievals and observations as an estimate of local retrieval error. The LiDAR snow depth data were aggregated to 90 m resolutions. Consistent with previous studies (Deems et al., 2013; Jacobs et al., 2021), LiDAR underestimation is more likely near forest edges due to partial occlusion and terrain effects. To mitigate the impact of measurement uncertainty LiDAR pixels with snow depths less than 20 cm are excluded from the evaluation. Additionally, high subgrid scale standard deviation of more than 0.3 typically found near the forest edges and highly heterogenous mixed pixels are removed from the final data (Singh et al., 2024)

*Snowpit Dataset*: Multiple snowpits excavated across Grand Mesa during the SnowEx'17 field campaign to obtain in situ measurements of snow water equivalent (SWE) (Table 1). While snowpit data along the SnowSAR flightlines on 21 February were limited, measurements collected between 20–24 February were included for evaluation. This assumes minimal changes in the snowpack over the four day period in the absence of significant snowfall event. Although localized variability may exist, broader spatial patterns such as the characteristic west-to-east gradient in snow depth are expected to remain stable.

## 3. Methodology

The methodology comprises four steps: i) Determination of $C_c$ and simulation of snowpack parameters using MSHM, ii) Determination of ground parameters, iii) Determination of vegetation parameters, and iv) Retrieval of snowpack parameters from SnowSAR backscatter. Figure 4 shows the flowchart for the proposed retrieval algorithm to retrieve the snowpack and vegetation parameters building on the Bayesian retrieval algorithm originally developed by Pan et al. (2023) and modified by Singh et al. (2024). The MSHM with forest shadow effect and interception parametrizations is used to simulate snowpack evolution since the beginning of the snow accumulation season for each pixel. For retrieval, the multilayered snowpack is transformed into a two-layered snowpack based on the relative change in the density profile criteria proposed by Singh et al.(2024). MEMLS is used to simulate snowpack and snow-ground backscatter, while the WCM is applied for vegetation backscatter. Table 2 provides the list of input and output variables associated with each model - MSHM-V, MEMLS-V, and the Bayesian RTM along with corresponding references.

**Table 2** - Models, input variables, and corresponding outputs used in the SWE retrieval framework

| Model | Input | Output | Reference |
|---|---|---|---|
| MSHM (Modified for Vegetation) | Rainfall<br>Temperature<br>Air Pressure<br>Incoming shortwave radiation<br>Incoming longwave radiation<br>Wind speed<br>Humidity<br>Albedo<br>LAI Based $C_c$ | Snow Temperature Profile<br>Soil Temperature Profile<br>Snow Density Profile<br>Snow Depth Layering Profile<br>Liquid Water Content Profile<br>Snow Correlation Length | MSHM: Cao and Barros (2020) |
| MEMLS + WCM | Snow Temperature Profile<br>Soil Temperature Profile<br>Snow Density Profile<br>Snow Depth Layering Profile<br>Snow Correlation Length Profile<br>Cross polarization fraction<br>Ground rms height<br>Frozen Vegetation Water Content | Diffused Reflectivity Profile<br>Specular Reflectivity Profile<br>Total Backscatter Coefficient | MEMLS: Proksch et al. (2015)<br>WCM: Bindlish and Barros (2001) |
| Informed Bayesian RTM for forested areas | Equivalent Snow Temperature Prior<br>Equivalent Soil Temperature Prior<br>Equivalent Snow Density Prior<br>Equivalent Snow Depth Prior<br>Correlation Length<br>Cross polarization fraction<br>Interpolated Ground rms height<br>Interpolated Frozen Soil Moisture<br>Frozen Vegetation Water Content<br>Total Backscatter Coefficient Prior<br>LAI Based $C_c$ | Optimized – Snow Layer Depth<br>Snow Density | Informed Bayesian RTM for Open areas: Singh et al., (2024) |

### 3.1. Numerical simulation of snowpacks using MSHM

The input datasets, including atmospheric forcing and remotely sensed variables were prepared and modified for integration into the modeling framework following Cao and Barros (2020). As previously discussed, canopy closure ($C_c$) was derived by downscaling the MODIS LAI product, removing zero values associated with open areas, and redistributing LAI values proportionally based on the GLAD tree height dataset. Incoming radiation and precipitation were adjusted using Eqs.(4-11). Specifically, incoming solar radiation was modified using canopy closure and transmissivity values obtained from literature, while precipitation was reduced based on interception calculated using the Hedstrom and Pomeroy (1998) model (HP98 hereafter). MSHM was further modified to incorporate intercepted snow unloading as an additional forcing input, with the corresponding snow density computed using Eqs. (12-13). The final processed datasets are used as inputs to inform and constrain the retrievals.

### 3.2. Determination of Ground parameters

The ground parameters were initially estimated following the methodology described in Singh et al., (2024), by setting the snow depth at 1 mm and using X-band HH-polarized backscatter in open areas. The retrieved ground parameter values showed minimal spatial variability, with changes not exceeding 1% from pixel to pixel, which allowed us to assume the validity of the First Law of Geography across different land cover types. This justified the use of ordinary kriging to spatially interpolate background priors into the forested regions. Nevertheless, some ground parameter values near forest edges appeared anomalously high due to significant residual errors in the simulated backscatter. To reduce the influence of these outliers, we restricted the interpolation to values within the 95% confidence interval, thereby improving the robustness and reliability of the estimated ground parameters.

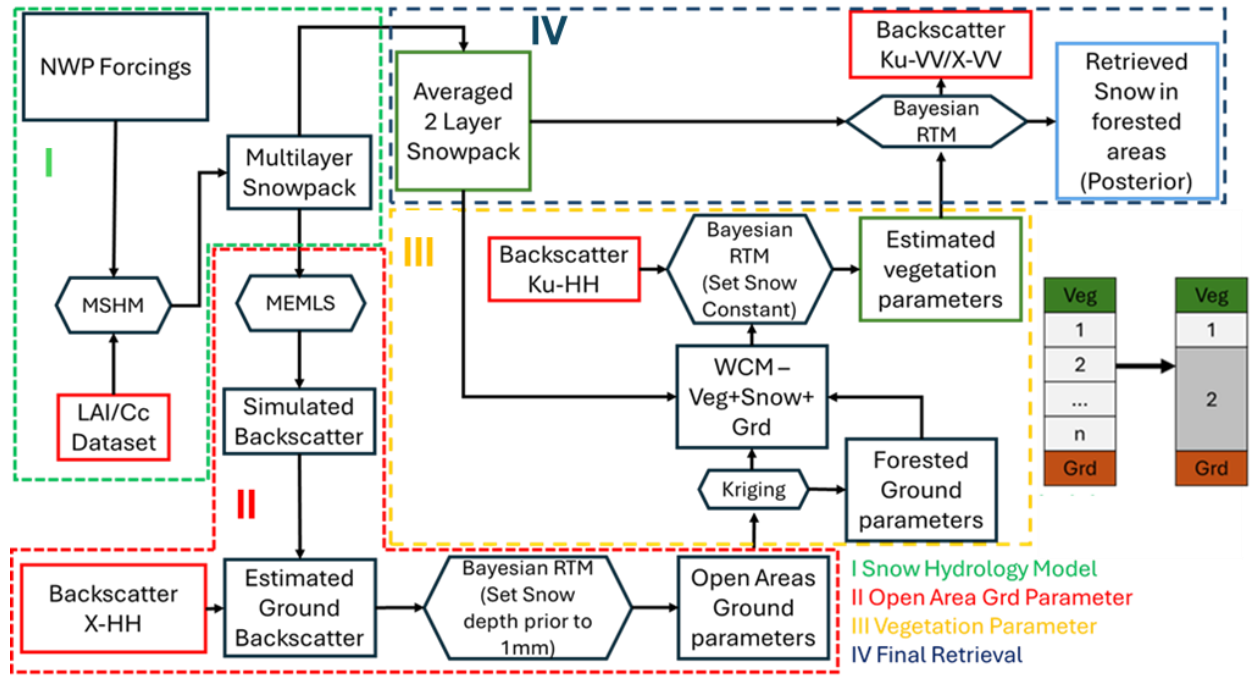

**Figure 4-** Methodology to determine SWE in forested environments. The workflow is divided into four main steps: (I) Snow Hydrology Model (green) – to simulate snow accumulation; (II) Open-Area Ground Parameterization (red) – to retrieve scattering and attenuation parameters for snow over in open areas; (III) Vegetation Parameterization (yellow) – to estimate vegetation-related parameters such as canopy transmissivity and vegetation water content; and (IV) Final Retrieval (blue) - integrates vegetation and ground contributions to estimate SWE within forested pixels.

### 3.3. Determination of vegetation parameters

To establish background and vegetation priors, we used Ku-band HH-polarized backscatter while holding snowpack parameters constant and allowing vegetation and ground parameters to vary. This initialization step provides the baseline for subsequent SWE retrieval. Given that all vegetation parameters (*A, B* and *Mv*) in the Water Cloud Model (WCM) are unknown, one parameter out of the three must be fixed to resolve the system and reach a unique solution. Note that in this study we rely on one-time measurements at each location as even in the case of overlapping pixels among SnowSAR flights there are significant differences in SAR geometry due to the low aircraft altitude. In the case of an operational satellite mission obtaining systematic measurements from long distances in Low Earth Orbit (LEO), it is possible to use the time-series measurements obtained at each location to infer (*A, B* and *Mv*) at characteristic timescales, and thus the parameter estimation problem would be highly simplified.

The apparent canopy architecture and water content will change principally due interception of snowfall in the cold season until the canopy structural storage capacity is exhausted, assuming weak winds and sublimation. Subsequent changes will be due to unloading and melting in the warm season that is not considered here. As the intercepted snowfall has low values, maximum value of 0.1 mm (Figure C.1), we omitted the backscatter from interception in our simulation. An initial value of $A = 0.0014$ is assumed as a starting point for optimization following Bindlish and Barros (2001). Since the WCM was originally developed for unfrozen conditions and only one parameter was fixed, all retrieved values are considered effective parameters rather than fully physical. During the final retrieval, ground surface roughness was held constant due to its high sensitivity, while soil moisture was treated as an uncertain parameter and optimized. Vegetation water content, retrieved during the initialization step, was assumed constant across frequencies and polarizations, and thus was not re-optimized in the second step (Table 5).

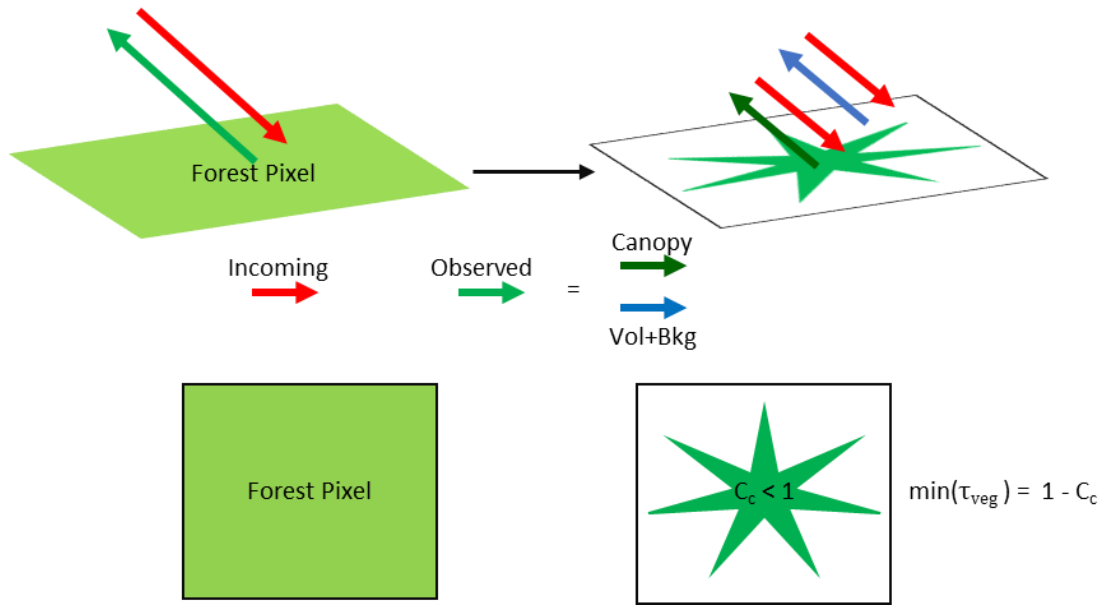

**Figure 5** - Conceptual illustration of radiative partitioning and microwave backscatter contributions in a mixed forest pixel. The left panel shows total observed backscatter from a pixel labeled "forest" in the land cover dataset, comprising both canopy-covered and open areas. The top panel separates this pixel into fractional components: canopy and open. Canopy closure ($C_c$) determines the fraction of vegetation-influenced backscatter. Minimum transmissivity is constrained by $1-C_c$, while maximum vegetation backscatter is bounded by assuming total reflection. These constraints are used to derive physical limits for Water Cloud Model (WCM) parameters as outlined in Eqs. (15–21).

Figure 5 shows the partitioning of radiation in a mixed pixel identified as vegetation or forest using the landcover dataset. Under an idealized partitioning scenario, the minimum transmissivity of vegetation within a mixed pixel is bounded by ($1-C_c$):

$$\min(\tau_{veg}) = 1 - C_c \tag{15}$$

Therefore Eq. (3) can be expressed as,

$$1 - C_c = e^{-2BMvsec\theta} \tag{16}$$

Similarly, the minimum observed backscatter is the combination of vegetation and open area backscatter:

$$\sigma_{obs} = \sigma_{veg} + (1 - C_c)\sigma_{vol+bkg} \qquad (17)$$

Keeping site specific A constant, $\sigma_{veg}$ depends on the vegetation water content. The maximum value of the vegetation water content, $Mv_{max}$ can be expressed as

$$Mv_{max} = \frac{\sigma_{obs} - (1 - Cc)\sigma_{vol+bkg}}{A\cos\theta} \qquad (20)$$

Similarly, in step IV of the retrieval algorithm (Figure 4), when we are optimizing A and B while keeping $Mv$ from step III constant, the maximum limits of the parameters are,

$$A_{max} = \frac{\sigma_{obs} - (1 - C_c)\sigma_{vol+bkg}}{Mv \cos\theta} \qquad (20)$$

$$B_{max} = -\frac{\log(1 - C_c)\cos\theta}{2Mv} \qquad (21)$$

These constraints provide a narrow and physically plausible range for the parameters improving the stability of the optimization scheme.

### 3.4. Retrieval of Snowpack Parameters

The multilayer snowpack predicted by the snow hydrology model (MSHM) is transformed into an equivalent two-layer snowpack following Singh et al. (2024) to prepare the snowpack priors for retrieval. The standard deviation and value ranges for snowpack, background and vegetation priors used in the optimization are detailed in Tables 3–6.

The final optimization simultaneously retrieves the following parameters: snow depth, snow density, snow correlation length, snow and soil temperature, frozen soil moisture, and vegetation backscatter coefficients (WCM parameters A and B), using dual-frequency X-VV and Ku-VV-backscatter. Figures B.1 and B.2 in Appendix show the spatial maps of final retrieved A and B for each frequency.

**Table 3 -** Ground parameter input mean, variance and range for the parameters. Frequency used to determine retrieve the parameter are mentioned in the Table. Alphanumeric subscript int – interpolated using open area parameters retrieved using Singh et al. (2024).

| Background Parameters | Frequency – Ku-HH | | | |
|---|---|---|---|---|
| | Mean | Variance | Range | |
| | | | Min | Max |
| Effective Soil Moisture, $Mvs$ | $Mvs_{int}$ | $0.3 \times Mvs_{int}$ | 0 | 1 |
| Ground Roughness, $GndSig$ | $GngSig_{int}$ | $0.3 \times GngSig_{int}$ | 0.0001 | 1 |

**Table 4 -** Vegetation parameter input mean, variance and range for the parameters used to optimize the Water Cloud Model in Step I (Figure 4)

| Vegetation Parameters | Frequency – Ku-HH | | | |
|---|---|---|---|---|
| | Mean | Variance | Range | |
| | | | Min | Max |
| Mv | $Mv_{avg} = \dfrac{Mv_{max} + Mv_{min}}{2}$ | $0.3 \times Mv_{avg}$ | $0.01 Mv_{max}$ | $\dfrac{\sigma_{obs} - (1-C_c)\sigma_{vol+grd}}{A\cos\theta}$ |
| B | $B_{avg} = \dfrac{B_{max} + B_{min}}{2}$ | $0.3 \times B_{avg}$ | $-\dfrac{\log(0.991)\cos\theta}{2Mv_{max}}$ | $-\dfrac{\log(1-C_c)\cos\theta}{2Mv_{max}}$ |

**Table 5 -** Vegetation parameter input mean, variance and range for the parameters used to optimize in Water Cloud Model in Step II (Figure 4).

| Veg Parameters | Frequency – X, Ku Band VV Pol | | | |
|---|---|---|---|---|
| | Mean | Variance | Range | |
| | | | Min | Max |
| A | $A_{avg} = \dfrac{A_{max} + A_{min}}{2}$ | $0.3 \times A_{avg}$ | $0.1 A_{max}$ | $\dfrac{\sigma_{obs} - (1-C_c)\sigma_{vol+grd}}{Mv\cos\theta}$ |
| B | $B_{avg} = \dfrac{B_{max} + B_{min}}{2}$ | $0.3 \times B_{avg}$ | $-\dfrac{\log(0.991)\cos\theta}{2Mv}$ | $-\dfrac{\log(1-C_c)\cos\theta}{2Mv}$ |

**Table 6** - Bayesian RTM model input variance and range for the priors derived from MSHM multilayer snowpack properties. The alphanumerical subscript in the 2-layer snowpack retrievals denotes layer number: 1- bottom layer; 2- top layer; avg is the average of all MSHM multilayer parameter values in the corresponding 2-layer snowpack.

| Snow + Background Parameters | Frequency – X-VV and Ku-VV | | | |
|---|---|---|---|---|
| | Variance, $\sigma^2$ | | Range for each layer | |
| | Bottom | Top | Min | Max |
| Snow Temp., Ts [°C] | $0.3 \times Ts_{1,avg}$ | $0.3 \times Ts_{2,avg}$ | $1.3 \times Ts_{min}$ | $0.7 \times Ts_{max}$ |
| Snow Density, $\rho$ [Kg/m$^3$] | $0.3 \times \rho_{1,avg}$ | $0.3 \times \rho_{2,avg}$ | $0.8 \times \rho_{min}$ | $1.2 \times \rho_{max}$ |
| Layer Snow Depth, DZ [m] | $0.1 \times DZ_1$ | $0.2 \times DZ_2$ | $0.2 \times DZ$ | $0.95 \times DZ$ |
| Correlation Length, $l_{ex}$ | $0.2 \times l_{ex,1,avg}$ | $0.2 \times l_{ex,2,avg}$ | $l_{ex,min}$ | $l_{ex,max}$ |

### 3.5. Evaluation of Retrievals

Point-scale validation was first conducted using independent SWE measurements from SnowEx'17 snow pits. For each pit, all SnowSAR pixels with centroids within a 100 m radius were identified, and the mean distance from the pit and the forest fraction within this buffered region were calculated using the reclassified land-cover dataset. Retrieval skill was quantified using mean absolute relative error (MARE). Successful retrievals were defined as pixels with local incidence angles between 30° and 50°, and relative residual backscatter (RRB) less than 30% (Singh et al., 2024).

The retrieved snow depth is compared against collocated LiDAR snow depth observations. LiDAR pixels with subgrid scale standard deviation of more than 0.3 m were removed due to high uncertainty. Metrics such as the root mean square difference (RMSD) and the Bhattacharya coefficient (BC) were used for assessing performance of the retrievals against LiDAR snow depth. BC is calculated using,

$$BC = \sum_{i=1}^{N} \sqrt{p_1(i)p_2(i)} \qquad (22)$$

Here $p_1$ and $p_2$ represent the normalized probability distributions of snow depth for the retrieved and reference (LiDAR) datasets, computed using 200 bins over a 3 m depth range. The BC quantifies the degree of overlap between these two distributions, where values close to 1 indicate strong similarity and good agreement (Bhattacharyya, 1942). To evaluate the impact of spatial averaging on retrieval accuracy, we also performed retrievals at the 30 m resolution for the shorter flight running perpendicular to the plateau. This allowed for a direct comparison with 30 m LiDAR data to assess potential errors introduced by upscaling of categorical landcover datasets. The methodology for generating and comparing high-resolution retrievals is consistent with the criteria applied at 90 m. As LiDAR snow depth is a modeled product, we report retrieval-LiDAR differences rather than errors proper.

## 4. Results

Figure 6 shows the spatial distribution of retrievals for all 4 flights and pit observations for dates between 21-23 Feb. Note the agreement between the increasing snow depth from west to east consistent with the pit dataset in Fig. 6a. The map of absolute relative errors in Fig. 6b shows small values below 20% except for the one pit on the edge of the plateau where the slope is very steep. To evaluate the retrievals against point-scale snow pit measurements and use as many pits as possible, all forested pixels within 100 m of the pits were considered.

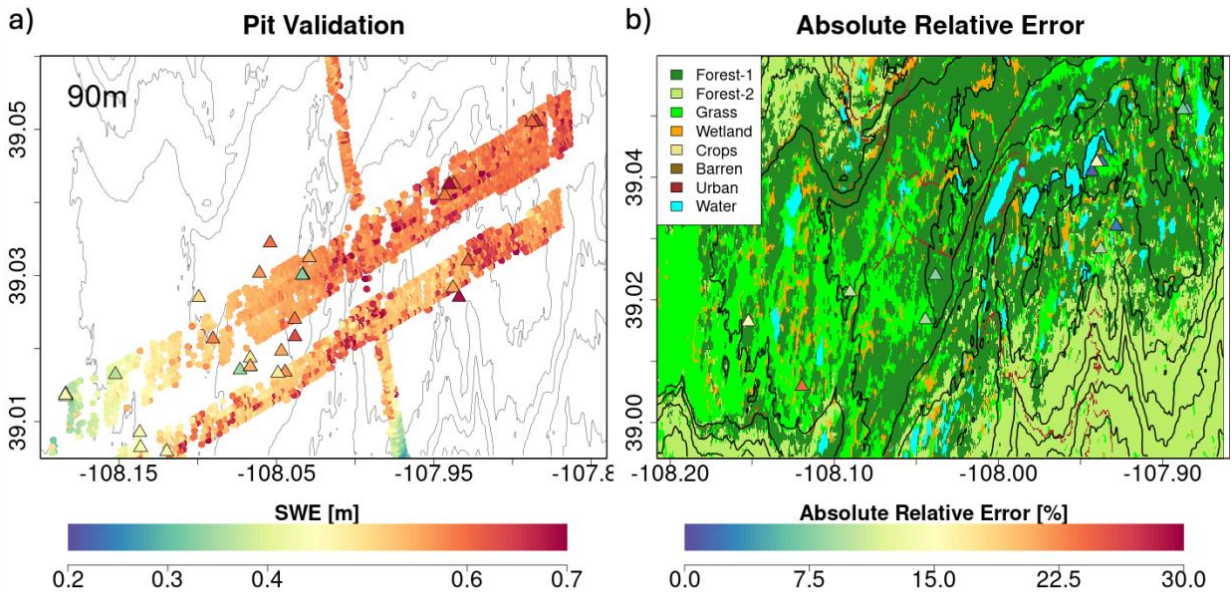

**Figure 6-** a) Comparison between retrieved SWE and pit datas. Successful retrievals are for pixels with local incidence angles in the 30º- 50º range and relative residual backscatter (RRB) of less than 30% for each of the four flights. b) Spatial distribution of pit locations marked by triangles overlaying a land cover basemap. The triangles are colored according to absolute relative error. . It should be noted that in the first map we are showing all pits whereas in the relative error is calculated for pits with atleast 1 retrieved pixel within 90 m.

Table 7 shows the summary evaluation of retrieved SWE against SWE from snowpit measurements. The overall RMSE is 0.033 m, that is 8% of maximum SWE. Pit32S with large mean distance from the pixels and low forest fractions (forest edge) showing high MARE of 25%.

**Table 7** - Evaluation of successful SWE retrievals at 90 m resolution against SWE at SnowEx'17 snow pits and retrieved snowpacks at 90 m resolution. All N pixels with centroids within 100 m of each snow pit snow depth. Standard deviation (SD) of all pixels within 100 m is calculated. Average distance from all pits (ADP) within 100 m is calculated. Average Pit Forest fraction (APFF) for the circle with 100 m radius around the pit is calculated

| x | y | Swe-Pit [m] | Swe-Ret [m] | SD (Ret) | MARE [%] | N pixels | ADP [m] | APFF | Date | Pit ID |
|---|---|---|---|---|---|---|---|---|---|---|
| -107.94 | 39.04 | 0.60 | 0.59 | 0.008 | 2 | 6 | 10 | 0.97 | 2/21/2017 | Pit71E |
| -107.93 | 39.03 | 0.61 | 0.59 | 0.047 | 3 | 3 | 27 | 0.21 | 2/21/2017 | Pit74E |
| -108.15 | 39.02 | 0.36 | 0.41 | 0.008 | 14 | 4 | 62 | 0.64 | 2/22/2017 | Pit26W |
| -108.12 | 39.01 | 0.44 | 0.55 | - | 25 | 1 | 71 | 0.39 | 2/22/2017 | Pit32S |
| -108.09 | 39.02 | 0.58 | 0.52 | 0.013 | 10 | 3 | 14 | 0.5 | 2/22/2017 | Pit38E |
| -108.04 | 39.02 | 0.57 | 0.52 | 0.001 | 9 | 2 | 45 | 0.58 | 2/22/2017 | Pit63E |
| -108.04 | 39.02 | 0.59 | 0.54 | 0.013 | 8 | 2 | 3 | 0.58 | 2/22/2017 | Pit66N |
| -107.89 | 39.05 | 0.66 | 0.62 | 0.01 | 6 | 2 | 30 | 0.64 | 2/22/2017 | Pit92E |
| -107.89 | 39.05 | 0.60 | 0.65 | - | 8 | 1 | 73 | 0.63 | 2/22/2017 | Pit92W |
| -107.94 | 39.04 | 0.59 | 0.59 | 0.016 | 0 | 6 | 6 | 0.97 | 2/23/2017 | Pit69N |
| -107.94 | 39.04 | 0.69 | 0.59 | 0.013 | 14 | 3 | 14 | 0.97 | 2/23/2017 | Pit71W |
| -107.94 | 39.03 | 0.56 | 0.61 | - | 9 | 1 | 22 | 0.74 | 2/23/2017 | Pit72S |

Figures 7 and 8 show the heatmaps of LiDAR snow depth against MSHM priors and SnowSAR retrieved snow depth, respectively. Histograms of the results with bin resolution of 0.015 m are shown in Fig. 9. The MSHM predicted priors do not capture the standard deviation for deeper snowpacks as expected due to underestimation of snowfall and other sources of uncertainty in operational forecasts (Cao and Barros, 2023). By contrast, the retrieved snow depths significantly improved the results both in terms of the range of snow depth and the SD of the distribution along the flight paths, with improved BC metric relative to the MSHM indicating that the spatial patterns of retrieved snow depth are in better agreement with the LiDAR.

Table 8 provides a summary of the intercomparison among the three snow depth datasets. The snow depth in the forest pixels varies within a narrow range of snow depths (1.5-1.75 m) for the LIDAR pixels, and there is a good agreement between the MSHM and the LiDAR snow depths. The performance of the retrieval algorithm deteriorates for the final shorter flight at 20:23:38 as illustrated by the misalignment of the flight histograms in Fig. 9 and the lowest BC metric in Table 8. As we can see in the heatmaps and in the snow depth histograms (Fig. 9) that there are some outliers in the retrieved backscatter retrievals with large standard deviation and increased RMSD, biasing the metrics. Informed Bayesian retrievals show improvement in both the standard deviation and BC for all flights.

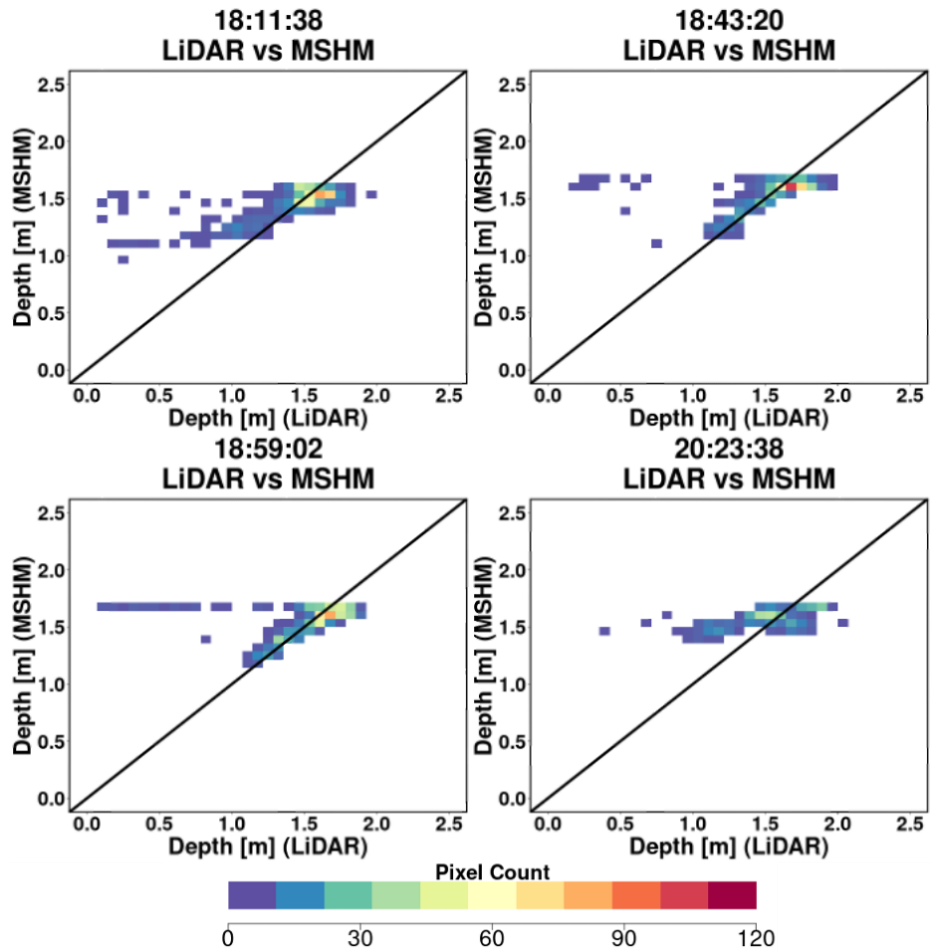

**Figure 7** – Heatmaps of LiDAR versus MSHM predicted snow depth for each SnowSAR flight at 90 m resolution for overlapping forest pixels only.

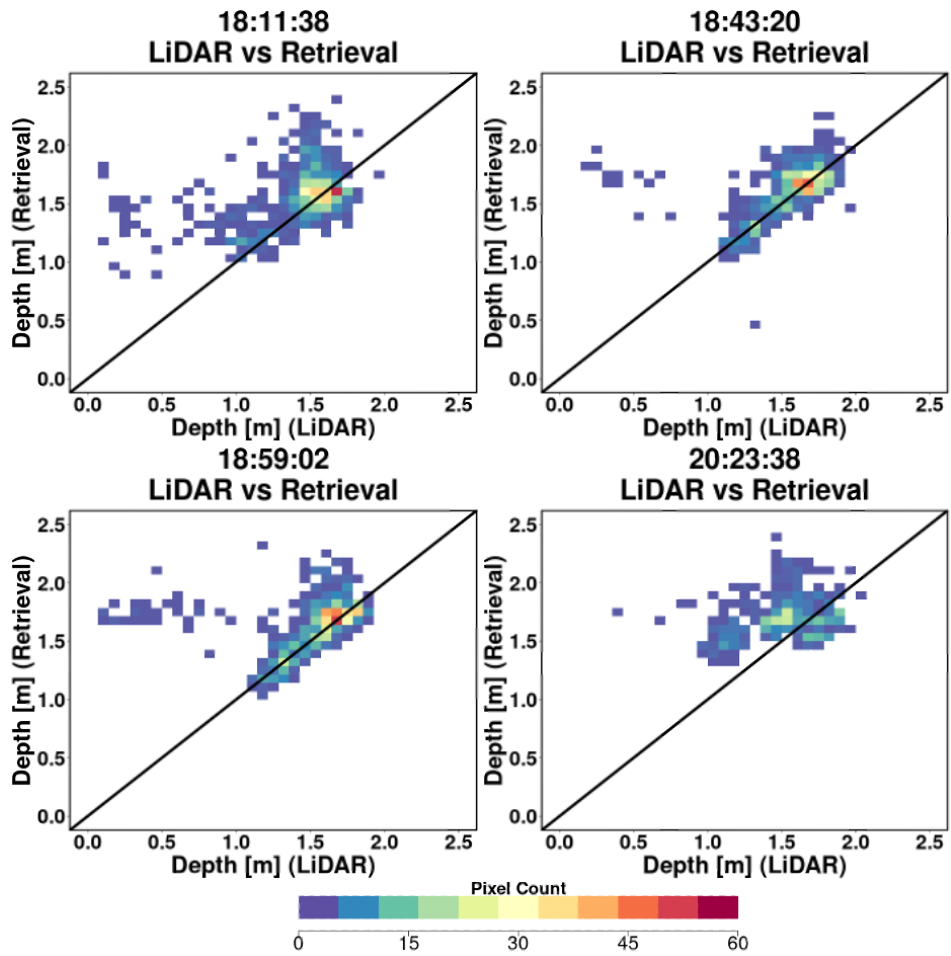

**Figure 8** – a) Heatmaps of LiDAR versus successful snow depth retrievals at 90 m resolution for forest pixels only. Successful retrievals are for pixels with local SnowSAR incidence angles in the 30º- 50º range and relative residual backscatter (RRB) of less than 30%.

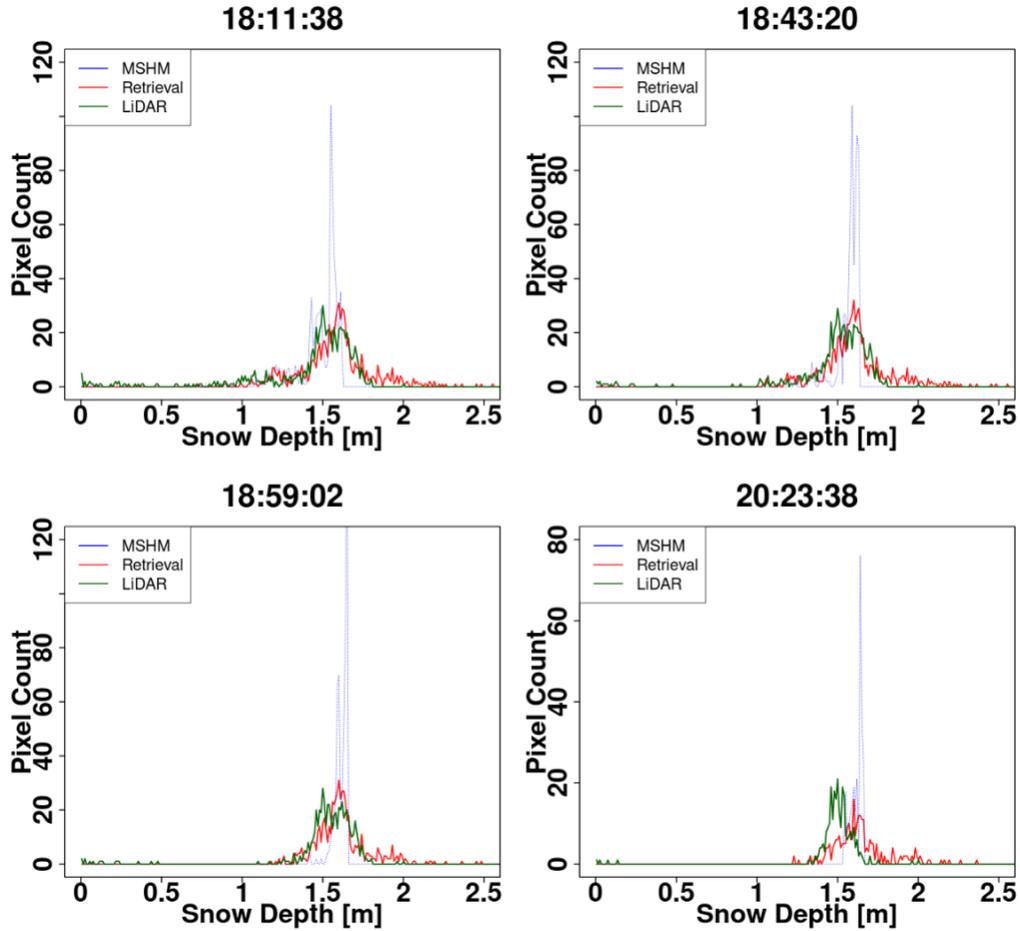

**Figure 9-** Histograms of snow depth predicted by MSHM, estimated from LiDAR measurements and for successful retrievals at 90 m using 2- layer snowpacks. The total number of pixels for each snow depth product is the same. Successful retrievals are for pixels with local incidence angles in the 30°- 50° range and relative residual backscatter (RRB) of less than 30% for each of the four flights. LiDAR snow depth in pixels with subgrid scale variability corresponding to standard deviation of less than 0.3 m for the upscaled 90 m LiDAR pixel are not included.

**Table 8** - Summary of snow depth statistics and error metrics at 90 m resolution: estimates from LiDAR measurements, MSHM predictions, and successful SnowSAR retrievals for forested pixels and subgrid-scale standard deviations of less than 0.3 m for the upscaled LiDAR pixel (Singh et al., 2024). BC – Bhattacharya coefficient (Eq. 15). Successful retrievals are for pixels with local incidence angles in the 30°- 50° range and relative residual backscatter (RRB) of less than 30% for each of the four flights.

| Flight | Mean [m] | | | Standard Deviation [SD] | | | BC | | RMSD | |
|---|---|---|---|---|---|---|---|---|---|---|
| | Retrieval | MSHM | LiDAR | Retrieval | MSHM | LiDAR | Ret-LiD | MSHM-LiD | Ret-LiD | MSHM-LiD |
| 18:11:38 | 1.59 | 1.50 | 1.46 | 0.22 | 0.10 | 0.30 | 0.87 | 0.76 | 0.15 | 0.06 |
| 18:43:20 | 1.62 | 1.55 | 1.56 | 0.20 | 0.12 | 0.24 | 0.94 | 0.70 | 0.06 | 0.01 |
| 18:59:02 | 1.64 | 1.55 | 1.53 | 0.20 | 0.12 | 0.30 | 0.89 | 0.78 | 0.11 | 0.03 |
| 20:23:38 | 1.71 | 1.58 | 1.51 | 0.20 | 0.08 | 0.27 | 0.78 | 0.68 | 0.2 | 0.07 |

Examination of the spatial maps of LiDAR (Fig. 10) and retrieved snow depth (Fig. 11) and absolute relative difference with respect to LiDAR snow depth (Fig. 12) enables identification of potential error sources. In the first three flights LiDAR estimates along the edges of open areas are attributed to upscaling errors in land cover category, that is the mixed pixel artifact. In the final flight (20:23:38), large errors are present on the northernmost slope and in the highly heterogeneous areas surrounding small ponds. As we are upscaling landcover using nearest neighbor interpolation, areas with contrasting land-cover types and thus dielectric heterogeneity contain signal from non-forest pixels thus introducing errors.

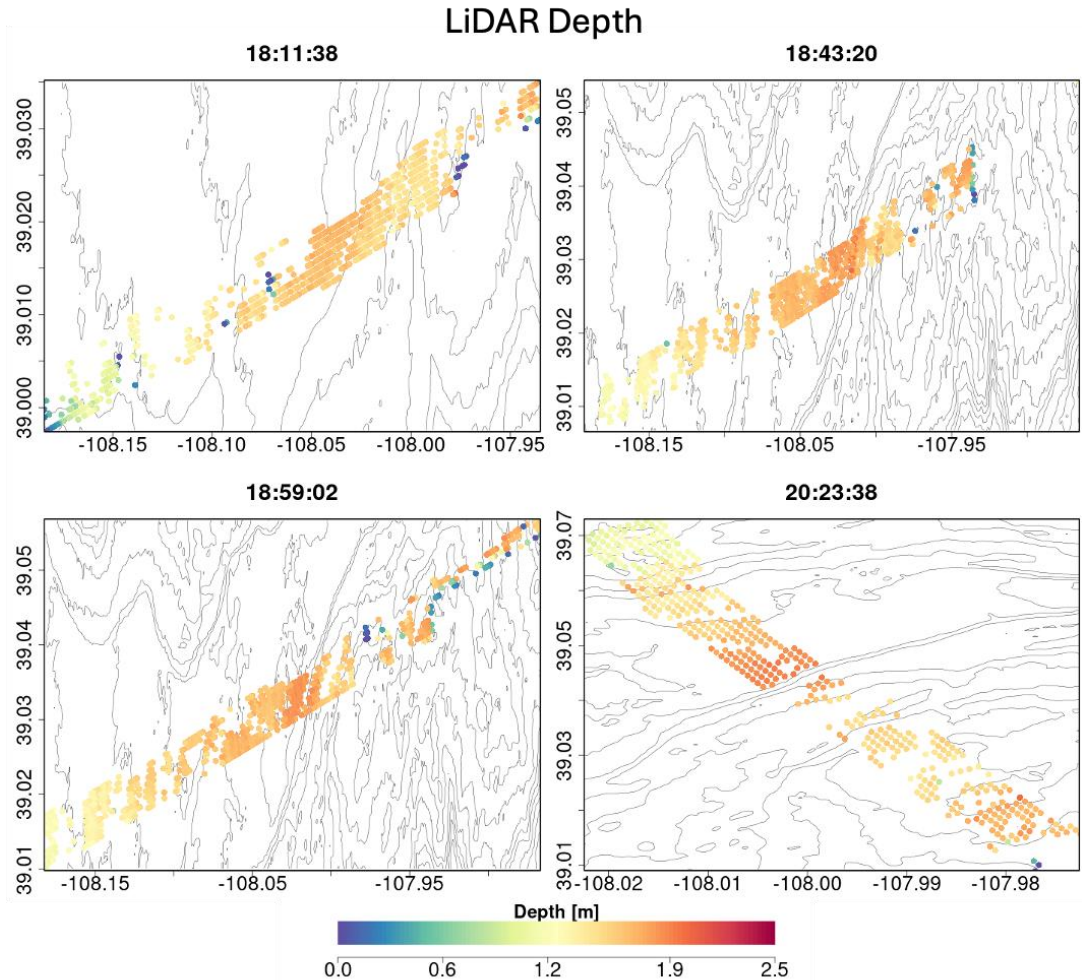

**Figure 10 -** Spatial distribution of LiDAR snow depth for forested areas with subgrid-scale standard deviation (SSD) of less than 0.3 m at 90 m resolution.

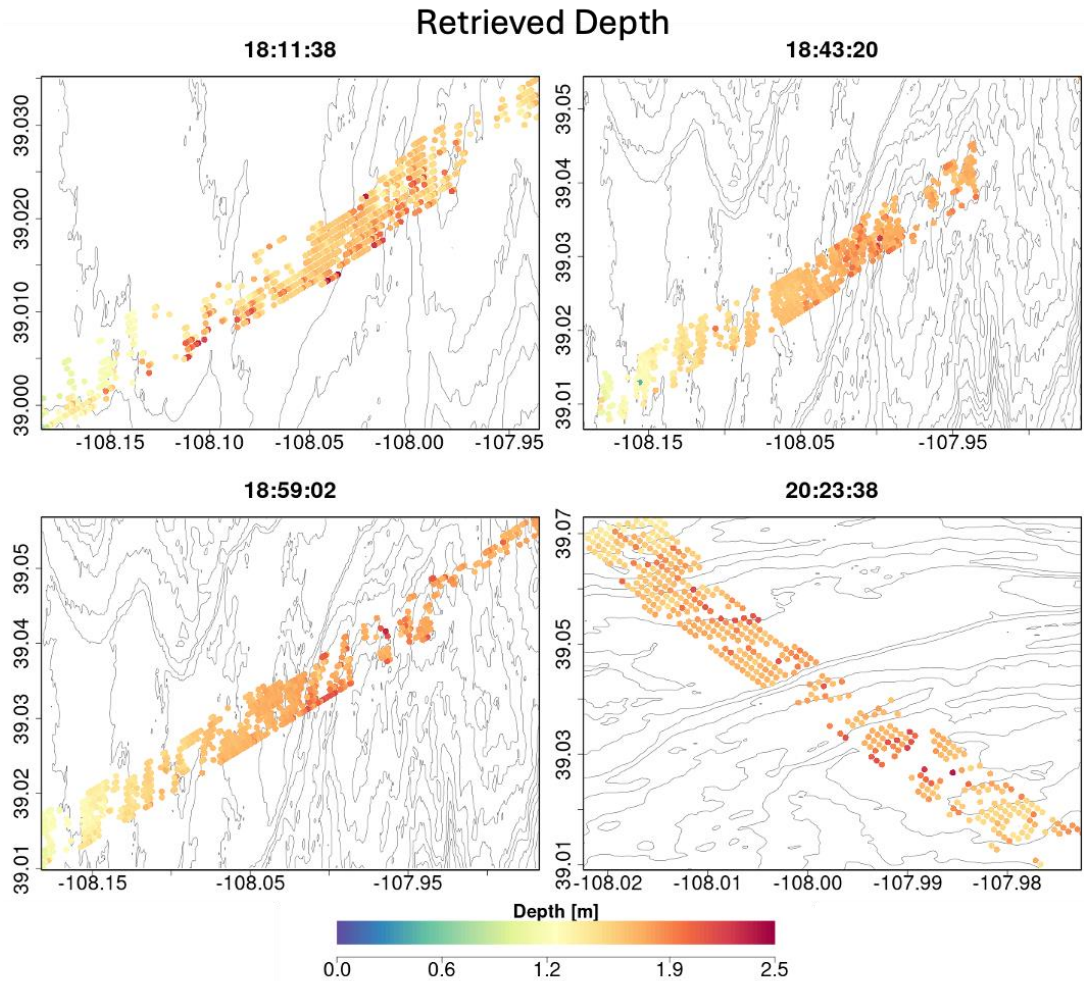

**Figure 11 -** Spatial distribution of snow depth retrievals at 90 m resolution for forested areas and for pixels with subgrid-scale standard deviation (SSD) of less than 0.3 m for the upscaled collocated LiDAR pixel.

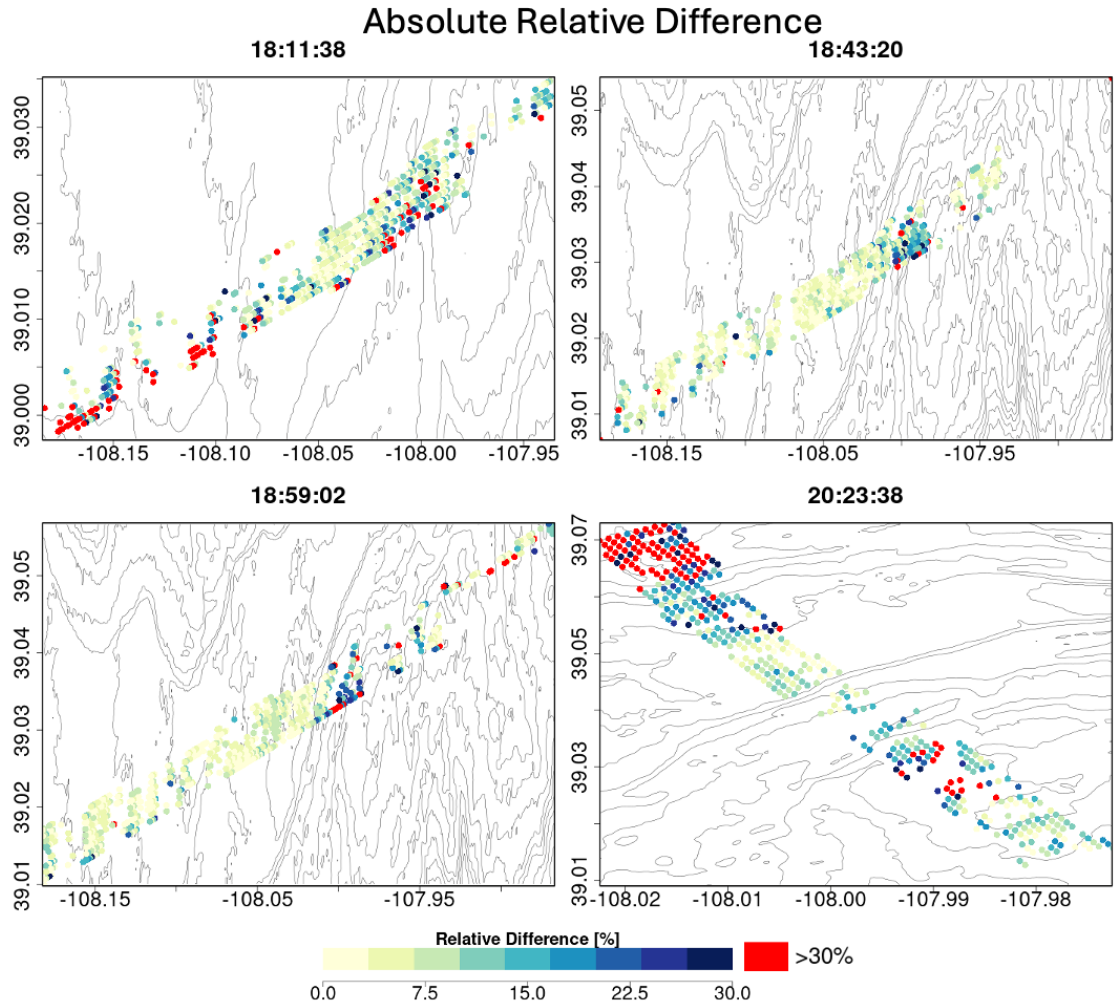

**Figure 12-** Absolute residual between snow depth from retrievals and LiDAR in Figures 10 and 11. Most of the high values are are either along the edge of the forest and grassland or in the mixed forest-pond-wetland pixels and on the slopes in the northern edge of the plateau.

As discussed in Section 3.5, to assess the errors introduced by the upscaling of categorical datasets such as forest cover and lake fraction, we calculated the subgrid-scale standard scale fraction of input parameters. Figure 13 shows the spatial distribution of forest fraction, water fraction, and canopy closure within a 90 m pixel. For flights 18:11:30 and 18:43:20, regions exhibiting large retrieval errors correspond closely with areas of high terrain variability, particularly along slope edges. In pixels with elevated forest or frozen lake fractions, biases in the backscatter signal due to mixed pixel effects propagate to the final SWE retrievals. Additional errors may arise from LiDAR underestimation of snow depth in sloped or densely forested areas as shown by Jacobs et al. (2021) and May et al. (2025). Jacobs et al. (2021) reported up to a 75% reduction in LiDAR point cloud density under dense canopy, leading to significant underestimation. Errors over steep terrain are further compounded by irregular ground point spacing, resulting in large interpolation errors (Deems et al., 2013). Therefore, the value of evaluating retrievals against LiDAR-based estimates in such conditions, especially in the fourth flight (20:23:38), is limited. Additionally, biases may also stem from the coarse resolution of atmospheric forcing datasets, such as HRRR precipitation (Cao and Barros, 2023), which may not capture microclimatic variations along forested slopes and thus introduce larger uncertainty in the priors (English et al., 2021; James et al., 2022).

Beyond the empirical nature of the parameterization of $C_c$, uncertainties in MODIS-derived LAI may also affect retrieval accuracy (Peng et al., 2024). Finally, potential instrument-related biases, including calibration and viewing geometry effects, remain an important area for future investigation. Nevertheless, we highlight the general retrieval success for pixels with forest fraction greater than 70%, meeting desired requirements (NASEM, 2018).

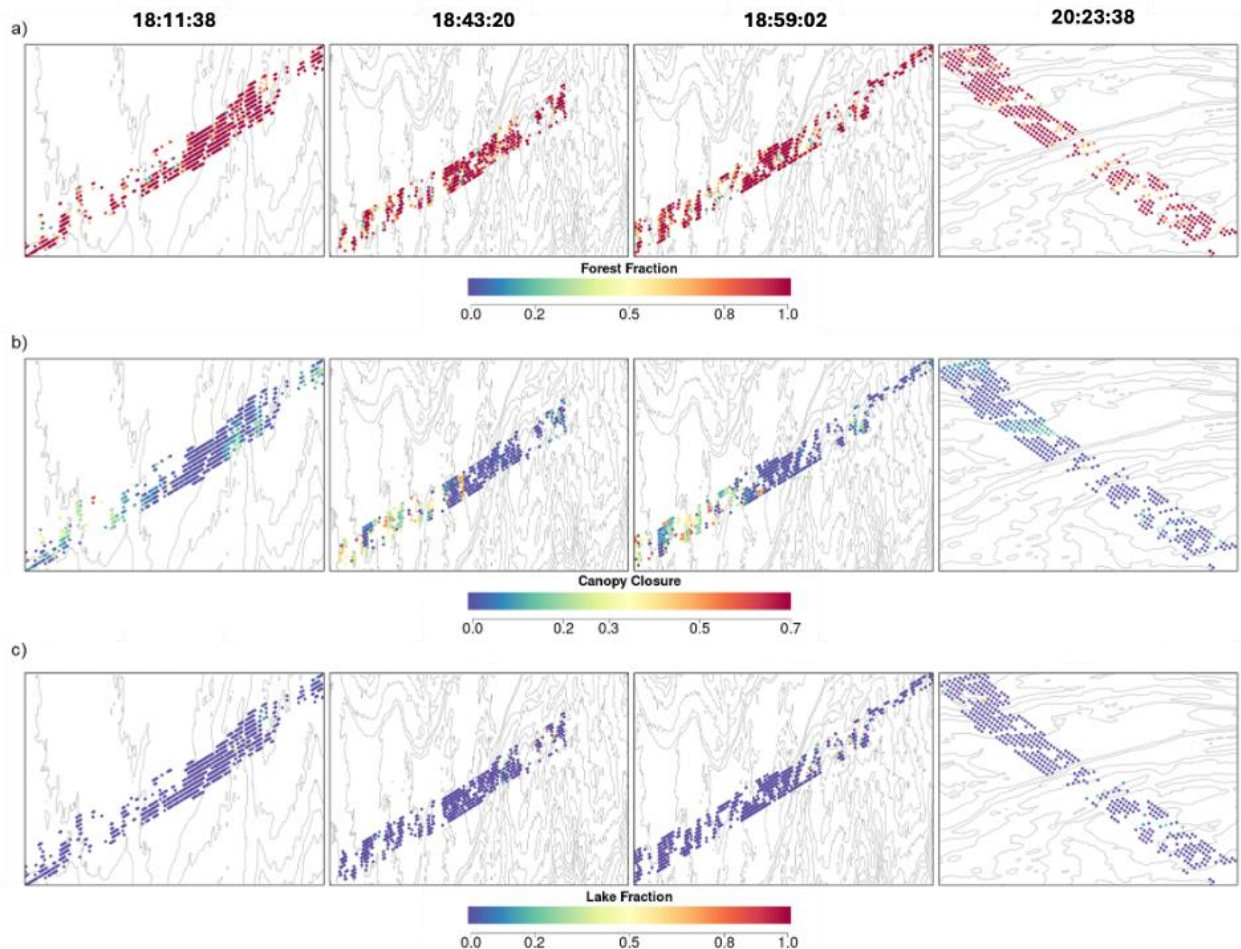

**Figure 13** – Demonstration of subgrid-scale variability of all pixels that impacts retrieval uncertainty: a) Forest fraction for a 90 m pixel using the 30 m land cover dataset. Improper upscaling of categorical datasets may introduce retrieval errors. b) Canopy closure uncertainty due to empirical estimation in heterogenous areas. c) Lake fraction for a 90 m pixel using 30 m land cover dataset. Difference between frozen water and snow dielectric properties may introduce a bias in the final retrievals.

To determine the errors introduced due to improper upscaling of the land-cover categories, the retrievals were repeated at 30 m resolution for the shorter SnowSAR flight (20:23:38). Figure 14 shows a comparison between the 30 m retrievals and the LiDAR data at the same resolution. The RMSD for the 30 m for retrieval compared to LiDAR is 0.2 (compared to 0.33 for 90 m retrievals) and BC of 0.87 (0.78 for 90 m). The 30 m retrievals show improved agreement with LiDAR in terms of both mean and variance, and the areal extent of large relative errors (> 30%) is significantly smaller than at 90m resolution. However, large errors remain in the northern edge of the flight on the forested slopes with high lake fraction. Additional analysis is imperative to eliminate the ambiguities needed to quantify the uncertainty in y both the SnowSAR backscatter measurements and LiDAR-based snow depth estimates in complex terrain.

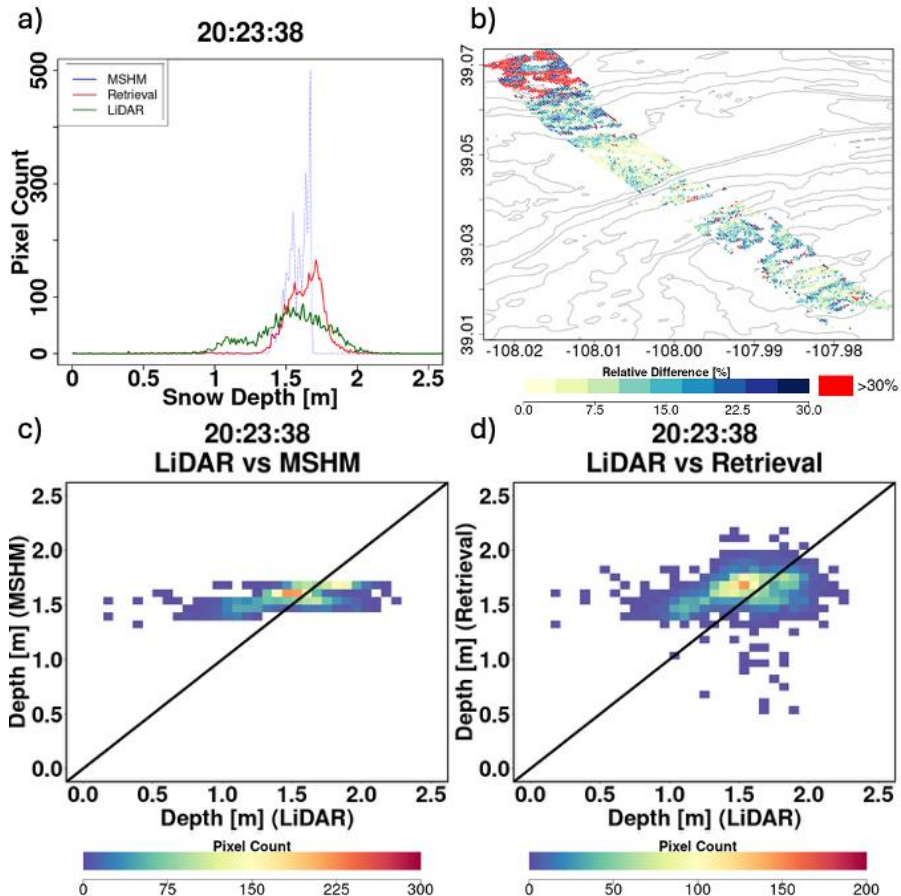

**Figure 14-** Comparison of 30 m SnowSAR retrievals and LiDAR snow depth for the 20:23:38 flight path. a) Retrieved snow depth at 30 m resolution. b) Absolute residual between SnowSAR retrievals and LiDAR snow depth. c) LiDAR snow depth at 30 m resolution and d) retrieved snow depth at 30 m. While the 30 m retrievals better capture mean and variance compared to coarser resolution retrievals at 90m, large residuals persist along the northern slope of the plateau due to terrain complexity and land cover heterogeneity (high lake faction).

## 5. Conclusion

A Bayesian coupled atmosphere-statistical framework, building on prior studies (Singh et al., 2024), was used to retrieve snow water equivalent (SWE) from airborne X- and Ku-band radar observations in the presence of trees. The approach produced consistent and reliable results across several SnowSAR flightlines over the forested areas in Grand Mesa, Colorado. Prior estimates of snowpack properties were generated using a multilayer snow hydrology model (MSHM) driven by atmospheric forcing from operational numerical weather prediction (NWP) forecasts and analyses. MSHM is modified to account for forest canopy shadow and interception effects. The multilayered snowpack was averaged to two layers to reduce the number of parameters to be optimized as in Singh et al. (2024). Retrievals were conducted independently for each 90 m pixel along SnowSAR flight tracks using VV-polarized backscatter, and constrained by prior distributions of snow, background and vegetation parameters derived from MSHM, MEMLS and the Water Cloud Model (WCM) to parameterize vegetation, respectively.

Following Singh et al., 2024 a 30% threshold for relative residual backscatter was selected as a balance between physical quality and data retention. As shown in Appendix Figure D.1 most pixels fall below 10%, but at 90 meter-scale SnowSAR observations show elevated variability near forest edges (where most pits are located), especially in Ku-band. Stricter limits would remove these physically meaningful signals and greatly reduce spatial coverage. The 30% criteria therefore preserve continuity across snow–vegetation transitions while remaining consistent with physical expectations. Note that the need to constrain the incidence angle only arises because of the geometry of the low altitude airborne measurements and it would go away for satellite-based SAR operations. The posterior distributions of retrieved snow depth were evaluated against collocated airborne LiDAR and snowpit SWE and snow depth measurements collected during the SnowEx'17 campaign, with validation restricted to pixels located within 100 m of pit locations. The retrievals show strong agreement with pit observations, achieving an RMSE of 0.033 m, or approximately 8% of the maximum SWE measured. Comparisons with ASO LiDAR snow depth estimates confirm that the SnowSAR retrievals improve the spatial distribution of snow depth and distribution of snowpack structure, particularly in regions with deeper snow accumulation.

Errors in retrievals were spatially associated with heterogeneous terrain and complex land cover transitions. Particularly, large uncertainties are observed near forest edges and steep slopes due to subpixel variability in elevation, forest fraction, and water presence. Validation performance in the fourth flight was limited due to increased terrain-induced noise and fewer valid pixels. Additional uncertainty arises from LiDAR underestimation in dense canopies, coarse resolution of NWP weather forcing especially precipitation, and empirical uncertainties in MODIS LAI-derived canopy closure. These sources of uncertainty highlight the importance of improving subgrid-scale land cover characterization, input resolution, and ancillary data quality in future applications. Specifically, errors in canopy closure estimates can be reduced using time series backscatter observations, particularly within the Ku-band region of the electromagnetic spectrum.

This study demonstrates the viability of a dual-frequency, physically informed Bayesian retrieval framework for SWE estimation in complex forested landscapes. With continued advances in high-resolution SAR observations and global atmospheric reanalysis products, the methodology is extensible to large-scale operational applications and is particularly well suited for satellite-based snow monitoring missions.

## 6. Appendix

### A. Determination of high-resolution Canopy Closure

Leaf Area Index (LAI) values derived from MODIS (MOD15A2H) at 500 m resolution were downscaled to 90 m to support high-resolution retrievals. MOD15A2H provides both the mean and standard deviation of LAI per pixel. Figure A.1 illustrates the conceptual steps taken to estimate canopy closure ($C_c$) for a given pixel. For each 500 m pixel, subgrid (30 m) non-forest pixels were identified and assigned a LAI value of zero. The forest-only LAI was then estimated by,

$$LAI_{forest} = LAI \cdot \frac{N_{all}}{N_{forest}} \tag{A..1}$$

Here $N_{all}$ is all subgrid 30 m pixels, $N_{forest}$ is the number of forest pixels and $LAI$ is leaf area index of the mixed pixel. The adjusted standard deviation, after excluding non-forest contributions, was calculated as,

$$\sigma_{forest} = \sqrt{\frac{\sigma_{all}^2 \cdot N_{all} - LAI^2 \cdot N_{open}}{N_{forest} - 1}} \tag{A.2}$$

Here $\sigma_{all}$ is the original standard deviation of LAI for the mixed pixel and $N_{open}$ is the number of open (non-forest) pixels. To spatially redistribute LAI across the 30 m forested subpixels, a normal distribution with the updated mean and standard deviation was sampled, using tree height as a covariate. The minimum and maximum bounds were defined as:

$$LAI_{min} = LAI_{forest} - 3\sigma_{forest} \tag{A.3}$$

$$LAI_{max} = LAI_{forest} + 3\sigma_{forest} \tag{A.4}$$

Several studies have reported a positive relationship between canopy height and LAI across forest types, indicating that taller vegetation generally supports greater leaf area (Yuan et al., 2013; Urrego et al., 2021). This relationship provides a physical basis for using vegetation height as a covariate in redistributing LAI within mixed pixels. Therefore, the final downscaled LAI for each 30 m pixel was assigned based on the local vegetation height (*VH*) using a linear interpolation:

$$LAI_{30m} = LAI_{min} + \left(\frac{VH - VH_{min}}{VH_{max} - VH_{min}}\right) \cdot (LAI_{max} - LAI_{min}) \tag{A.5}$$

This approach preserves subgrid variability and accounts for canopy structure in estimating canopy closure, a critical input for SWE retrievals in forested environments.

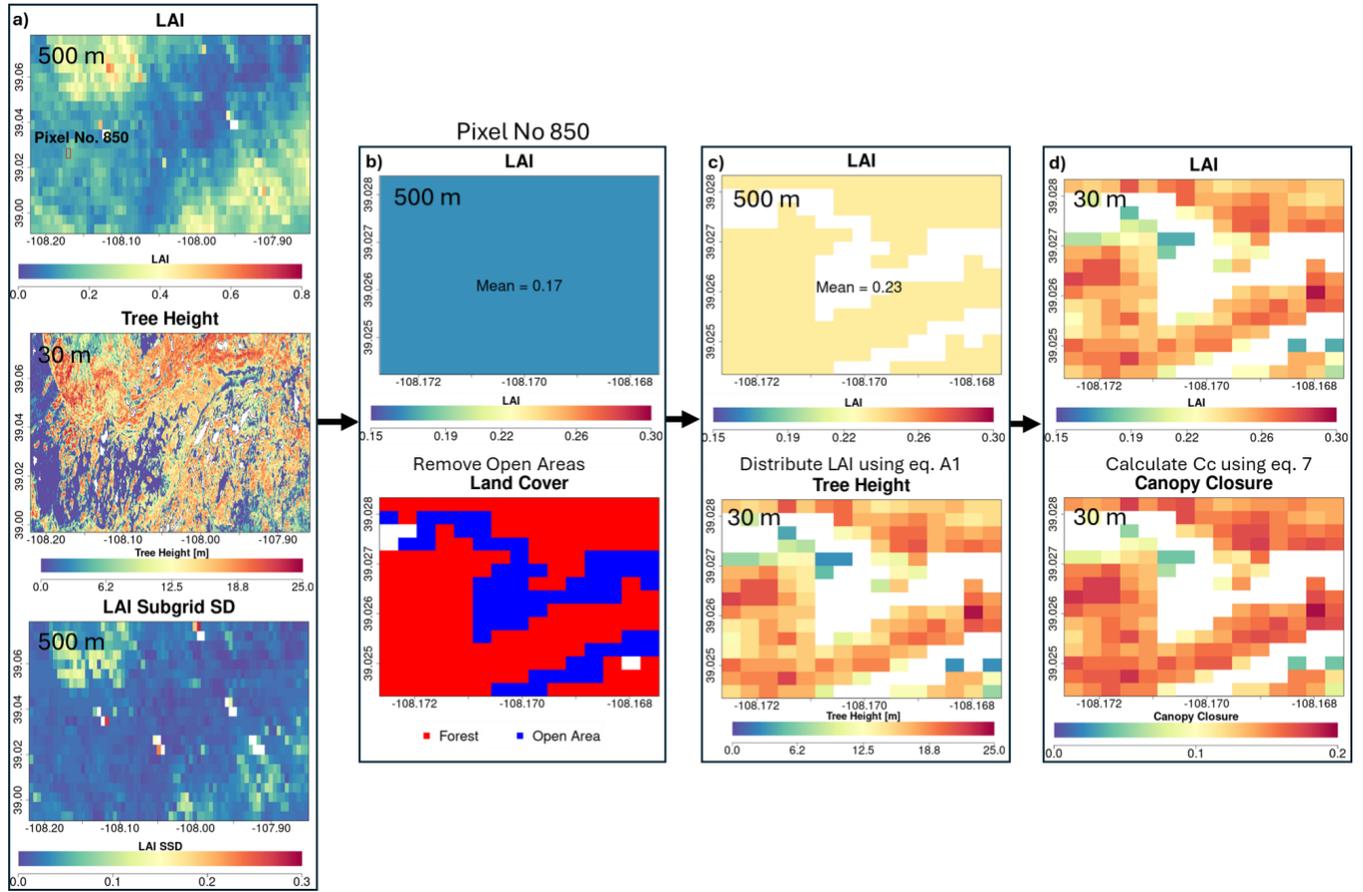

**Figure A.1** – Showing the schematics downscaling of one 500 m MODIS LAI pixel to 30 m Canopy closure dataset. a) MODIS LAI dataset at 500 m, tree height dataset at 30 m, subgrid scale standard deviation (SSD) for 500 m MODIS pixel. b) LAI and land cover distribution Pixel No. 850. c) adjusted LAI after removing open areas. d) Redistribution and sampling of LAI using adjusted mean/standard deviation and tree height as spatial covariate.

## B. Retrieved Vegetation Parameters

WCM parameters were retrieved after final step for all available frequencies and polarizations. Spatial distributions of the retrieved parameters *A* and *B* are presented in Figures B.1 and B.2, respectively. Figure B.3 illustrates the comparison of retrievals for pixels common to two different flight lines acquired at distinct viewing angles. Overall, the retrieved parameters show strong agreement between flights, with the exception of parameter *A* for the X-band VV polarization, which exhibits greater variability. Figure B.4 presents the spatial distribution and cross-flight comparison of the retrieved vegetation water content ($M_v$), further demonstrating consistency across overlapping observations. To further quantify cross-flight consistency, Figure B.5 shows the spatial distribution of absolute relative differences, for all retrieved vegetation parameters for all collocated pixels between the two flights. Differences are generally small (<10–20%) and spatially coherent along the flight track, increasing only near forest edges or terrain breaks where canopy heterogeneity is strongest. Poor agreement is present for A-Ku(VV) which needs to be studied further.

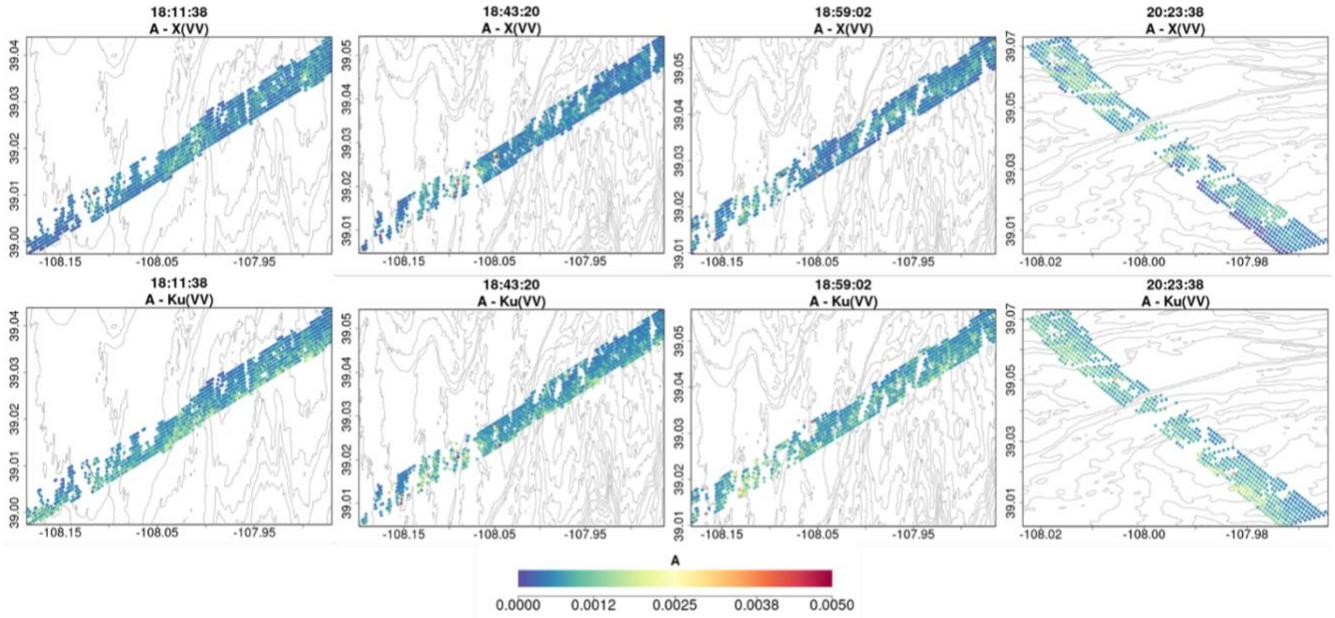

**Figure B.1** Spatial plot of Water cloud model parameter A for both channels (X and Ku VV) after final optimization run.

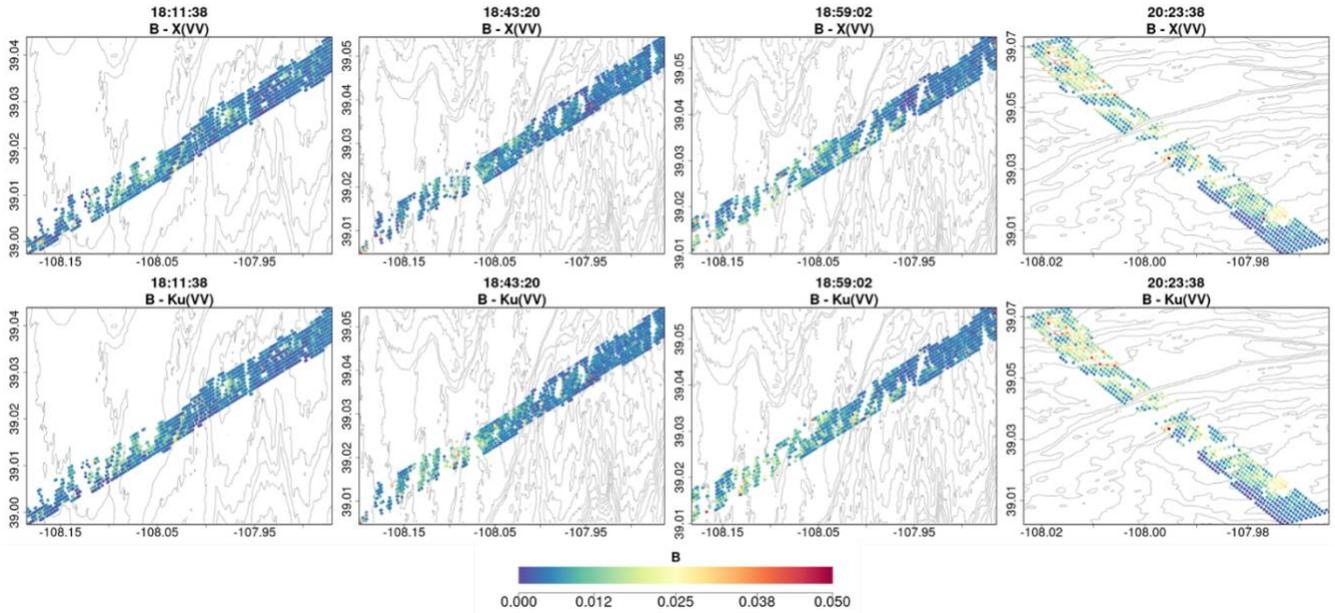

**Figure B.2** Spatial plot of Water cloud model parameter B for both channels (X and Ku VV) after final optimization run.

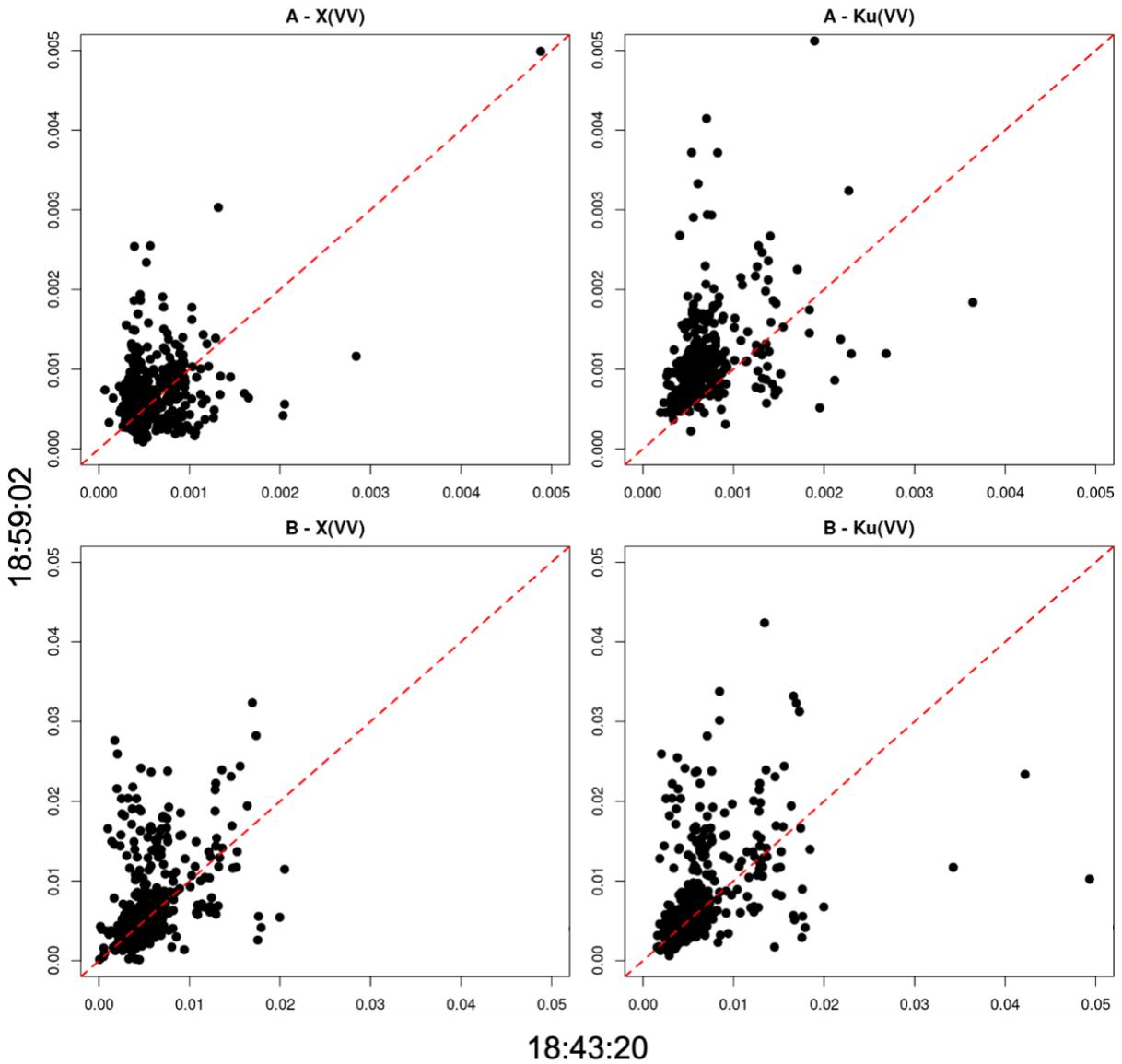

**Figure B.3** - Scatterplots of the retrieved A and B parameters from the Water Cloud Model (WCM) for overlapping pixels between the SnowSAR flights at **18:43:20** and **18:59:02** on 23 February 2017. The A parameter represents the canopy attenuation coefficient, while B represents the vegetation scattering term. Each panel shows results for X- and Ku-band VV polarization.

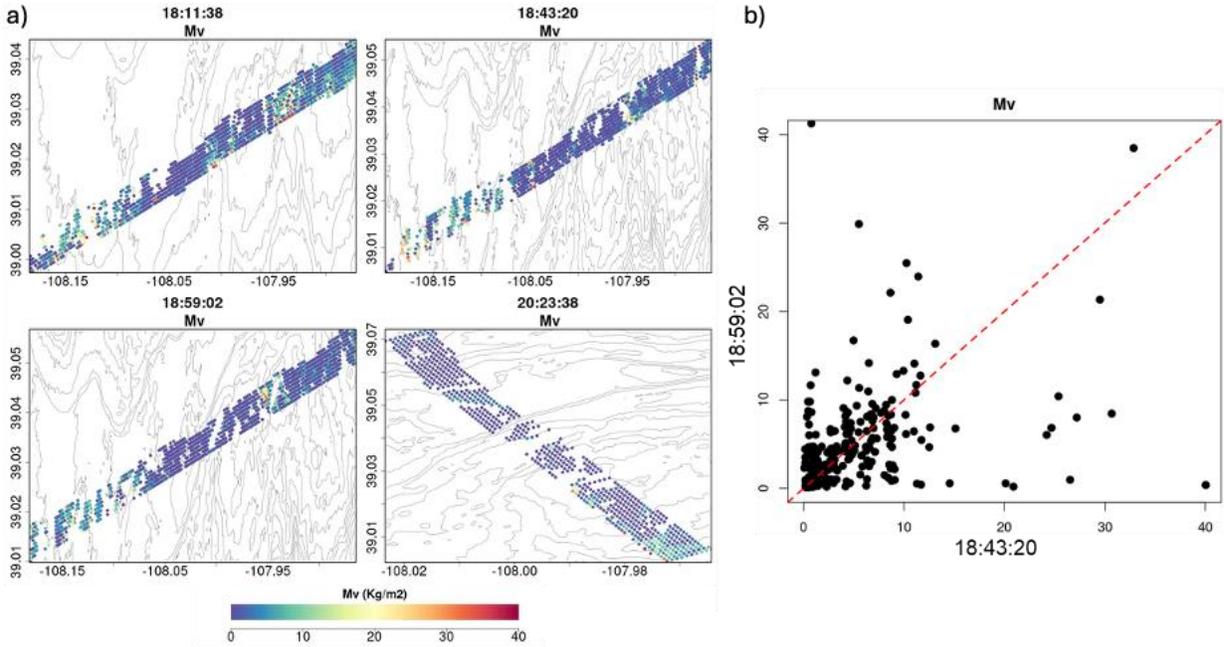

**Figure B.4** – a) Spatial distribution of retrieved vegetation water content (Mv, kg m$^{-2}$) from four consecutive SnowSAR flights on 23 February 2017 over Grand Mesa, Colorado, showing consistent flight-line coverage and temporal changes in canopy wetness. (b) Scatterplot of Mv for overlapping pixels between the 18:43:20 and 18:59:02 flights, illustrating cross-flight consistency of the retrieved vegetation water content.

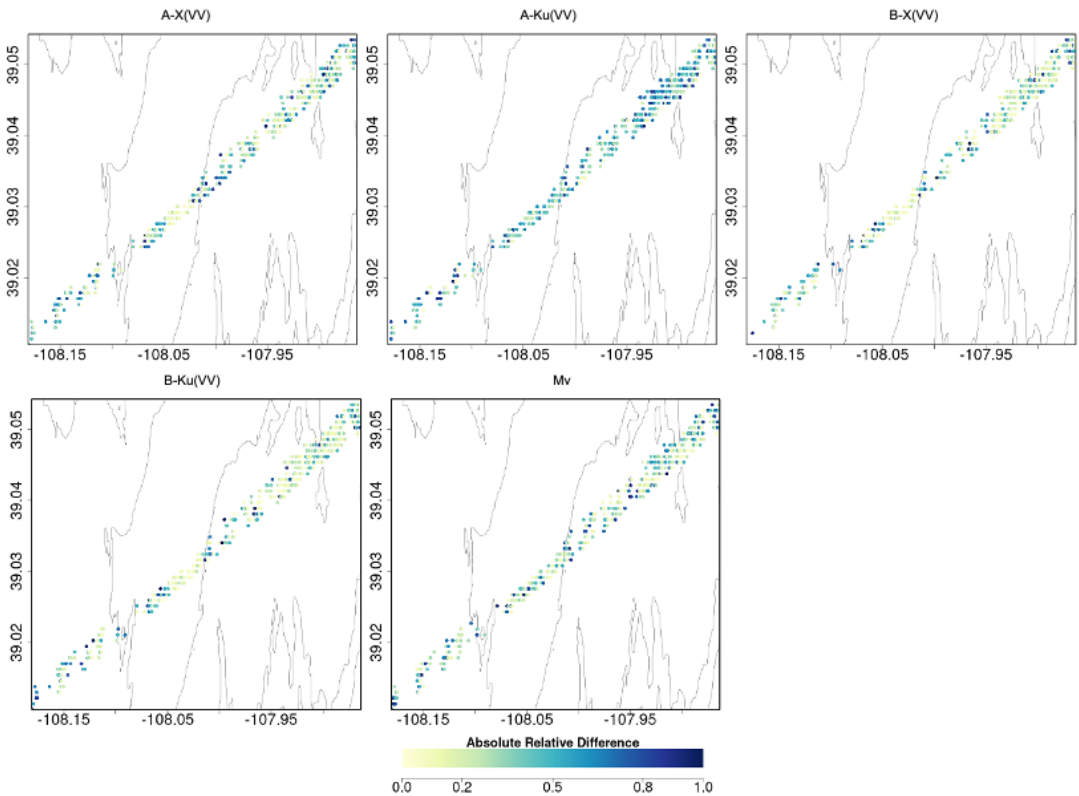

**Figure B.5** - Absolute relative differences between collocated pixels of the two consecutive SnowSAR flights (18:43:20 and 18:59:02) for A-X, A-Ku, B-X, B-Ku, and Mv.

## C. Interception

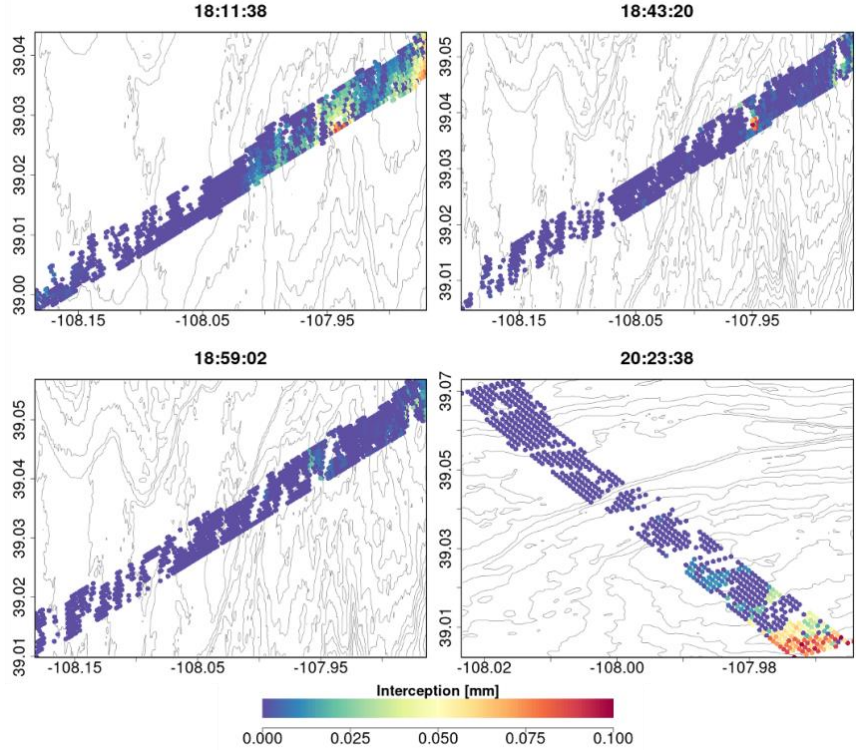

**Figure C.1** - Interception in mm during the flights calculated using HP 98. Areas with high canopy closure (Cc) show high interception value. However, the interception is too low to produce or affect the backscatter simulations.

## D. Residual Backscatter

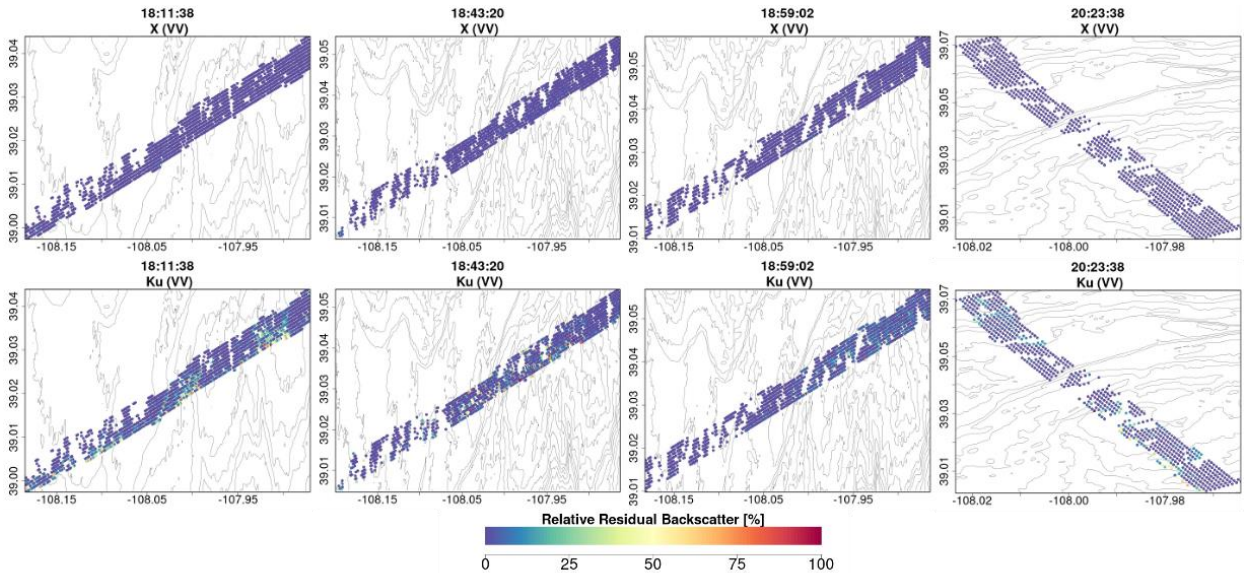

**Figure D.1 -** Relative residual backscatter (%) for X-band (top row) and Ku-band (bottom row) VV acquisitions at four overpass times. Colors indicate the magnitude of the normalized backscatter residual with respect to model predictions. Most pixels fall below ~10% error, while a limited number of higher-residual pixels appear primarily along forest edges in Ku-band, where vegetation structure drives greater sub-pixel variability. These results support the choice of a 30% residual threshold, which preserves physically meaningful signals.

## 7. Abbreviations

ADP- Average distance from all pits

APFF - Average Pit Forest Fraction

LAI – Leaf Area Index

DEM – Digital Elevation Model

RMSE – Root Mean Square Error

$R^2$ – Coefficient of Determination

MSE – Mean Squared Error

ARE – Absolute Relative Error

NALCMS – North American Land Change Monitoring System

SAR – Synthetic Aperture Radar

SnowSAR – Dual-frequency (X/Ku) airborne SAR instrument

MSHM – Multilayered Snow Hydrology Model

MEMLS – Microwave Emission Model of Layered Snowpacks

MEMLS-V – Vegetation-augmented MEMLS

WCM – Water Cloud Model

RTM – Radiative Transfer Model

Base-AM – Bayesian Active Microwave retrieval framework

MCMC – Markov Chain Monte Carlo

HRRR – High-Resolution Rapid Refresh atmospheric model

NWP – Numerical Weather Prediction

LiDAR – Light Detection and Ranging

IEM – Integral Equation Model (microwave scattering)

RRB – Relative Residual Backscatter

BC – Bhattacharyya Coefficient

NLDAS – North American Land Data Assimilation System

GLAD – Global Land Analysis & Discovery tree-height dataset

Ku-band – Ku-band radar frequency (~13–18 GHz)

X-band – X-band radar frequency (~8–12 GHz)

HH, VV, HV – Radar polarization channels (horizontal/vertical)

**Symbols**

$\sigma_{total}$ – Total Reflectance within a mixed pixel

$\sigma_{vol}$ – Volume backscatter from within the snowpack

$\sigma_{bkg}$ – Background backscatter at the snow–ground interface

$\sigma_{veg}$ – Vegetation backscatter

$T_s$ – Volume backscatter from within the snowpack

$\rho$ – Snow Density

$Dz$ – Snow depth

$Q$ - Cross-polarization fraction used in MEMLS/Base-AM

$C_c$ - Canopy-closure fraction

$A, B$ - WCM parameters

$Mv$ – Vegetation Water Content

$SW$ - Shortwave Radiation

$SW$ - Longwave Radiation

$\epsilon_v$ - Vegetation Emmissivity

$I$ - Interception

$P_i$ - Precipitation

$f_u$ – Wind Unloading

$f_m$ – Melt Unloading

$l_{ex}$ – Correlation Length

$\tau$ - Transmissivity

## 8. Acknowledgements

The authors gratefully acknowledge the insightful feedback provided by the Barros Research Group at the University of Illinois. Additionally, the authors extend their appreciation to the dedicated scientists, engineers, and technicians at NASA, USGS, ASO, Finnish Met. Institute and Canada Meteorological Service, for their hard work in making valuable satellite datasets freely available.

## 9. Author contributions

APB and SS conceptualized the study. SS developed and implemented the retrieval framework, including modifications and coupling of the codes, under the guidance of APB. SS completed the retrievals and analyzed the results under the guidance of APB and CV. SS, APB and CV wrote and performed the intial review on the paper and will reply to the reviewers.

## 10. Financial support

This research has been supported by the National Aeronautics and Space Administration (grant no. NNX17AL44G to Ana P. Barros) and the National Oceanic and Atmospheric Administration Cooperative Institute for Research to Operations in Hydrology (CIROH) agreement (grant no. NA22NWS4320003 to Ana P. Barros).

**12. Competing interests**. The contact author has declared that none of the authors has any competing interests.

## 13. Declaration of generative AI and AI-assisted technologies in the writing process

Statement: During the preparation of this work SS used Grammarly and ChatGPT to correct the grammar, subject verb agreement, voice (Active and Passive) and tense. Additionally, SS used the tools to change the reference 'Chicago' style to the recommended formal. After using this tool/service, the authors reviewed and edited the content as needed and take full responsibility for the content of the published article.